# Asteroid phase curves using sparse *Gaia* DR2 data and differential dense light curves

E. Wilawer ,[1]⋆ D. Oszkiewicz,[1] A. Kryszczyńska,[1] A. Marciniak,[1] V. Shevchenko,[2] I. Belskaya,[2]
T. Kwiatkowski,[1] P. Kankiewicz ,[3] J. Horbowicz,[1] V. Kudak,[4] P. Kulczak,[1] V. Perig[4] and K. Sobkowiak[1]

[1]*Astronomical Observatory Institute, Faculty of Physics, Adam Mickiewicz University, ul. Słoneczna 36, 60-286 Poznan, Poland*
[2]*Department of Astronomy and Space Informatics, V. N. Karazin Kharkiv National University, 4 Svobody Sq., Kharkiv 61022, Ukraine*
[3]*Institute of Physics, Jan Kochanowski University, ul. Uniwersytecka 7, 25-406 Kielce, Poland*
[4]*Laboratory of Space Researches, Uzhhorod National University, Daleka st. 2a, 88000, Uzhhorod, Ukraine*

## ABSTRACT

The amount of sparse asteroid photometry being gathered by both space- and ground-based surveys is growing exponentially. This large volume of data poses a computational challenge owing to both the large amount of information to be processed and the new methods needed to combine data from different sources (e.g. obtained by different techniques, in different bands, and having different random and systematic errors). The main goal of this work is to develop an algorithm capable of merging sparse and dense data sets, both relative and differential, in preparation for asteroid observations originating from, for example, *Gaia*, *TESS*, ATLAS, LSST, *K2*, VISTA, and many other sources. We present a novel method to obtain asteroid phase curves by combining sparse photometry and differential ground-based photometry. In the traditional approach, the latter cannot be used for phase curves. Merging those two data types allows for the extraction of phase-curve information for a growing number of objects. Our method is validated for 26 sample asteroids observed by the *Gaia* mission.

## 1 INTRODUCTION

Asteroid phase curves describe the dependence of asteroid brightness on the geometry of scattering determined by the phase angle, namely the angle between the directions to the observer and the Sun as seen from the asteroid. Phase curves are of general interest because they allow us (1) to probe the surface properties (such as regolith particle size, surface roughness, porosity) of asteroids; (2) compute the absolute magnitude, which is related to the asteroid size and albedo; and (3) determine various physical properties of asteroids, for example the taxonomic type (Shevchenko et al. 2016; Oszkiewicz et al. 2021).

Typical asteroid magnitude phase curves are linear in the phase angle range 8°–40° and become non-linear at larger angles, where they are influenced by surface roughness on the topographic scale. At small phase angles, the so-called opposition effect is observed, which appears as a non-linear increase of the asteroid brightness towards opposition. According to our current knowledge, the opposition effect is formed by coherent back-scattering and shadow-hiding effects, the linear part is determined by the shadow-hiding effect and the contribution of single-particle scattering, and at large phase angles a contribution of the shadow-hiding effect appears at all roughness scales (e.g. Muinonen et al. 2002). Shadow hiding is caused by single scattering, and the effect is more pronounced for dark surfaces and less evident for bright surfaces, where multiple scattering is more dominant. The coherent back-scattering mechanism dominates for high-albedo surfaces. This explains the strong dependence of phase-curve behaviour on the albedo of the asteroid (Belskaya & Shevchenko 2000). The linear phase slope can be used for a reliable estimation of the albedo of asteroids from phase-curve measurements (Belskaya & Shevchenko 2018; Belskaya & Shevchenko 2000; Oszkiewicz et al. 2021; Shevchenko et al. 2021). The phase-curve behaviour also depends on wavelength. For moderate- and high-albedo asteroids whose albedos increase with wavelength, so-called phase-reddening is observed (see Reddy et al. 2015).

Several advanced photometric models have been proposed to describe phase curves: Hapke's (Hapke 1963, 1966, 1981, 1984, 1986, 2002, 2008, 2012; Hapke & Wells 1981), Akimov's (Akimov 1975, 1979, 1988) and Shkuratov's (Shkuratov et al. 2011). However, owing to the limited accuracy of photometric observations, in most cases empirical phase functions with only two to three parameters are used to fit phase curves. The $H$, $G$ phase function was developed by Bowell et al. (1989) and adopted by the International Astronomical Union (IAU) in 1985. However, with increasing numbers of observations it became clear that the $H$, $G$ function overestimates the absolute magnitude of low-albedo objects and underestimates it for high-albedo objects. Therefore, two new $H$, $G_1$, $G_2$ and $H$, $G_{12}$ functions were proposed by Muinonen et al. (2010). The system was later improved by Penttilä et al. (2016). A new $H$, $G^*_{12}$ system was proposed to supersede the previous $H$, $G_{12}$ function. In the new model, one linear function of first degree is used to describe the dependence of $G_1$ and $G_2$, in contrast to the two functions used before. Furthermore, the use of one-parameter phase functions was proposed for objects with a small number of observations and/or low-quality photometry. In these functions, only the absolute magnitude $H$ is fitted, and the $G_1, G_2$ parameters are assumed for each taxonomic type, for example as in Shevchenko et al. (2016). For observations obtained at phase angles greater than 8° (such as those obtained by for example the *Gaia* mission), only the linear part of the phase curve can be fitted. The directional coefficient $\beta$ of the linear function is correlated with the geometric albedo (Belskaya & Shevchenko 2018).

⋆ E-mail: wilawerek@amu.edu.pl

Traditionally, phase curves are derived from relative photometry[1] dense light curves that are obtained from targeted ground-based observations. Currently, these high-quality phase curves are available for a few hundred asteroids, and the increase in the number of objects is small (Shevchenko et al. 2016). This is mostly due to the time-consuming nature of these observations and to the careful planning and calibration required to obtain them (Shevchenko 1997; Shevchenko et al. 2002, 2016; Pravec et al. 2012; Oszkiewicz et al. 2021). Those standard observations allow us to account for changes in brightness owing to aspect (if enough data are gathered and the shape can be determined), irregular shape, and rotation. The latter is often referred to as the light-curve amplitude correction (LC-AC hereafter). Owing to the fact that light-curve morphology changes with phase angle and the minima are usually more affected than the maxima, the maximum of light curves is typically used to construct phase curves (Shevchenko et al. 2002). On the other hand, phase curves derived from sparse photometry available from surveys are numerous but often burdened with large uncertainties. This is mainly caused by not accounting for the rotational brightness modulation (LC-AC) and manifests as a large spread of points around the fitted phase function.

The application of phase functions to large amounts of sparse photometry data was started a decade ago. By processing a large amount of photometric data from the Lowell Observatory photometric data base, Oszkiewicz et al. (2011) found a homogeneity of phase-curve parameters in asteroid families and studied the distribution of these parameters for various taxonomic classes (Oszkiewicz et al. 2012). Similar studies were performed by Vereš et al. (2015) using the Pan-STARRS data and by Waszczak et al. (2015) using the Palomar Transient Factory survey data. Recently, Mahlke, Carry & Denneau (2021) and Ďurech et al. (2020) confirmed the wavelength dependence of phase-curve coefficients using the sparse ATLAS dual-band photometry. That effect has been studied since the 1970s (Millis, Bowell & Thompson 1976; Miner & Young 1976; Gradie, Veverka & Buratti 1980; Gradie & Veverka 1986; Reddy et al. 2012; Sanchez et al. 2012). Alvarez-Candal et al. (2021a) derived phase curves in the Sloan Digital Sky Survey (SDSS) bands for about 12 000 asteroids, but did not study the wavelength dependence in detail.

Colazo & Duffard (2020) determined the $H, G$ parameters for about 4000 asteroids observed by the *Gaia* mission and available in the DR2 catalogue. They noted that the standard mean $G$ slope value ($G = 0.15$) currently used by the Minor Planet Center differs from the value observed for the *Gaia* data set. Yoshida et al. (2020) presented over 40 000 phase curves obtained from serendipitous observations of asteroids present in the data from the Tomo-e Gozen transient survey aimed at detecting young supernovae. Lin et al. (2020) fitted the $H, G$ phase function to about 1000 near-Earth asteroids observed by the Zwicky Transient Facility and assigned them a taxonomic type based on the $G$ slope value.

Currently, the accuracy of phase curves derived from sparse photometry is very limited, and we need new approaches to achieve better accuracy. Muinonen et al. (2020) presented a method for the simultaneous inversion of rotation periods, pole orientations, shapes, and phase-curve parameters using Bayesian statistics. The method was applied to three asteroids having numerous data in the *Gaia* DR2 catalogue. Other efforts are ongoing. Martikainen et al. (2021) combined the *Gaia* data with the sparse and dense-in-time photometry from the Database of Asteroid Models from Inversion Techniques (DAMIT) to perform convex inversion and ellipsoid inversion and derived the linear slopes for about 20 asteroids. Alvarez-Candal et al. (2021b) used Bayesian statistics to account for rotational modulation in the SDSS multi-filter photometry. Colazo, Duffard & Weidmann (2021) fitted the $H, G$ phase function to the *Gaia* DR2 data combined with relative photometry from the Asteroid Photometric Catalogue (Lagerkvist et al. 1995).

With the increasing amount of sparse photometric data, there is clearly a need for computationally efficient methods that can process large data volumes and at the same time reduce the uncertainty of the determined phase-curve parameters. This can be achieved by merging multiple measurements that are diverse in nature, which gives auxiliary knowledge unobtainable from sparse data alone. In this work, we present a new approach for obtaining asteroid phase curves from various information sources: combining the absolute but sparse *Gaia* data and data from ground-based differential photometry. Although straightforward, this approach has not yet been fully explored in the literature. In comparison to other works, it allows rotational brightness changes and complicated light-curve morphologies to be accounted for. The method presented here can be extended to other sparse photometric surveys.

In Section 2 we describe the data sources used in this study, and in Section 3 we consider the phase-curve fitting algorithm. The results are presented in Section 4, the discussion in Section 5, and a summary and plan for future work in Section 6.

## 2 DATA SOURCES

We use traditional differential ground-based photometric observations gathered at the Astronomical Observatory Institute of Adam Mickiewicz University in Poznań and absolute *Gaia* measurements. Combining these different data types (differential, dense-in-time photometry, and sparse but absolute photometry) allows us to account for the rotational component in the sparse photometry (through the inclusion of well-determined periods and higher than second-order Fourier series determined from ground-based differential photometry) and to derive the phase slope ($\beta$) parameter.

Routine asteroid photometric observations have been performed at Poznań Observatory since the 1990s, and thus decades of asteroid photometric observations are available. Howerver, for the purpose of this work, we select only observations obtained in the *Gaia* DR2 era to ensure similar aspect angles and viewing geometries; that is, we use observations obtained in the years 2014–2016. Furthermore, we use only observations of asteroids that have both differential ground-based photometry and *Gaia* measurements obtained for at least three different phase angles. The observing log for these objects is summarized in Table 1, in which we include rotational periods, aspect changes, time ranges of observations, the number of dense light curves, the number of Julian days of *Gaia* observations, and the number of transits registered by *Gaia* for each asteroid.

---

[1] By relative photometry we understand brightness measurements tied to a set of standard stars. Absolute photometry, as provided by *Gaia*, is tied to a laboratory source. Differential photometry results from the direct comparison of asteroid and stellar images present in the same field of view and is not tied to any particular magnitude scale.

**Table 1.** The columns present: the observed asteroid, the rotation period taken from Marciniak et al. (2015, 2018, 2019, 2021) and Shevchenko et al. (2021), the aspect change during the period of observation, the minimum phase angle observed by *Gaia*, the number of partial light curves obtained from ground-based telescopes, the number of Julian days of *Gaia* observations, and the number of transits registered by *Gaia*.

| Asteroid | Rotation period (h) | Aspect change (°) | Period of observations | $\alpha_{min}$ (°) | $N_{ground}$ | $N_{Gaia}$ | $N_{transit}$ |
|---|---|---|---|---|---|---|---|
| (70) Panopaea | 15.812 | – | 2014-08-14–2015-02-11 | 18.03 | 15 | 6 | 14 |
| (108) Hecuba | 14.255 | 7 | 2015-03-05–2015-12-20 | 12.58 | 10 | 9 | 15 |
| (109) Felicitas | 13.194 | 40 | 2015-05-01–2016-02-24 | 20.31 | 13 | 5 | 6 |
| (159) Aemilia | 24.493 | 9 | 2015-01-08–2015-10-23 | 12.51 | 11 | 9 | 27 |
| (195) Eurykleia | 16.527 | 6 | 2014-09-24–2015-05-30 | 15.1 | 8 | 6 | 9 |
| (202) Chryseis | 23.671 | 10 | 2014-08-30–2015-01-21 | 13.65 | 9 | 3 | 4 |
| (202) Chryseis | 23.666 | 9 | 2015-09-07–2016-04-04 | 18.75 | 10 | 5 | 8 |
| (260) Huberta | 8.2895 | 10 | 2014-09-12–2015-02-07 | 13.9 | 11 | 4 | 6 |
| (301) Bavaria | 12.240 | 17 | 2015-03-01–2015-12-24 | 16.8 | 9 | 10 | 14 |
| (305) Gordonia | 12.893 | 7 | 2014-08-08–2015-02-11 | 13.26 | 13 | 5 | 5 |
| (329) Svea | 22.777 | 15 | 2014-07-30–2015-02-06 | 16.82 | 11 | 3 | 4 |
| (362) Havnia | 16.923 | 10 | 2015-07-01–2016-02-28 | 21.2 | 6 | 6 | 8 |
| (380) Fiducia | 13.716 | 7 | 2015-07-02–2016-02-27 | 20.3 | 8 | 5 | 9 |
| (439) Ohio | 37.46 | 16 | 2014-08-21–2015-02-06 | 13.55 | 12 | 5 | 10 |
| (483) Seppina | 12.723 | 9 | 2014-09-17–2015-04-17 | 14.31 | 5 | 6 | 7 |
| (483) Seppina | 12.721 | 12 | 2015-11-09–2016-05-09 | 14.78 | 5 | 3 | 3 |
| (501) Urhixidur | 13.174 | 12 | 2014-09-11–2015-04-29 | 13.6 | 10 | 7 | 10 |
| (537) Pauly | 16.301 | 10 | 2015-11-15–2016-05-20 | 12.07 | 8 | 7 | 9 |
| (552) Sigelinde | 17.143 | 8 | 2015-05-04–2016-02-20 | 12.52 | 5 | 11 | 20 |
| (611) Valeria | 6.982 | 11 | 2014-08-29–2015-02-13 | 16.17 | 3 | 4 | 6 |
| (618) Elfriede | 14.799 | 12 | 2014-08-28–2015-01-21 | 14.09 | 12 | 3 | 6 |
| (618) Elfriede | 14.799 | 12 | 2015-09-06–2016-04-28 | 14.97 | 6 | 4 | 6 |
| (653) Berenike | 12.488 | 14 | 2015-02-23–2015-10-28 | 17.35 | 7 | 6 | 8 |
| (666) Desdemona | 14.617 | 7 | 2014-11-06–2015-08-19 | 13.34 | 18 | 9 | 14 |
| (666) Desdemona | 14.617 | 13 | 2016-01-14–2016-07-23 | 13.59 | 9 | 5 | 6 |
| (667) Denise | 12.683 | 15 | 2014-12-29–2015-10-25 | 11.14 | 6 | 9 | 18 |
| (667) Denise | 12.683 | 4 | 2016-03-08–2016-08-31 | 12.5 | 9 | 3 | 5 |
| (723) Hammonia | 5.4349 | – | 2014-08-08–2015-02-03 | 14.32 | 17 | 5 | 14 |
| (727) Nipponia | 5.0692 | – | 2014-10-25–2015-05-29 | 21.83 | 5 | 4 | 6 |
| (780) Armenia | 19.898 | 17 | 2015-02-20–2015-10-28 | 13.66 | 24 | 8 | 13 |
| (834) Burnhamia | 13.874 | 6 | 2014-09-20–2015-05-31 | 11.24 | 10 | 9 | 17 |
| (834) Burnhamia | 13.879 | 8 | 2015-11-15–2016-05-19 | 13.47 | 7 | 5 | 6 |

For all asteroids, we verified the aspect changes based on all available models (usually two) and reported the largest aspect change. The aspect changes calculated with ISAM (Interactive Service for Asteroid Models) (Marciniak et al. 2012) were less than 15° for most of the observed objects during the observation period. In many cases, the aspect change of the objects is smaller than the uncertainties in the spin-axis coordinates. Light curves and periods for all asteroids except for (611) Valeria and (727) Nipponia were taken from Marciniak et al. (2015, 2018, 2019, 2021) and Shevchenko et al. (2021). Observing circumstances for 611 and 727 are summarized in Table A1.

We also used sparse, but absolute *Gaia* photometry (Prusti et al. 2016). The *Gaia* mission provided photometric data of millimagnitude accuracy for about 14 000 asteroids (DR2) (Brown et al. 2018; Spoto et al. 2018). Most objects were observed at phase angles in the range $5° < \alpha < 30°$, and thus *Gaia* does not cover the opposition area. Typically, observations obtained during transits in two *Gaia* fields of view are separated by 106.5 min and thus can be reduced to two points on a light curve during a single rotation. Those two points are reduced to a single point on the phase curve, as the change of the phase angle between two *Gaia* transits is negligible (except for near-Earth objects). In the current catalogue, the total number of observations for a particular asteroid is very low. As a consequence, sparse data points alone are not enough to obtain reasonable-quality phase curves. Therefore, merging with other data sources and/or information is necessary. Here, we focus on merging diverse data sets, as indicated in Table 1.

From the *Gaia* catalogue, we extracted the following columns for our 26 objects: *number_mp*, indicating the asteroid number; *epoch_utc*, the observing time; and *g_mag*, the brightness in the *G* band and the corresponding flux *g_flux* with its uncertainty *g_flux_error*. A constant value of 2455197.5 d needs to be added to *epoch_utc*. The *G* magnitude error is estimated as:

$$g_{magerr} = -2.5 \log_{10} \left( \frac{g_{flux} + g_{fluxerr}}{g_{flux}} \right). \quad (1)$$

Each asteroid data are divided into oppositions based on solar elongation at the date of observation. One opposition comprises data from 0° to 360° solar elongation. Each asteroid opposition contains *Gaia* observations from at least three separate Julian days (typically there are two transits per Julian day) and at least one observing night from ground-based observatories covering all phases of the rotational period.

## 3 PHASE-CURVE FITTING ALGORITHM

The differential ground-based photometry is composed together with absolute *Gaia* measurements to determine the rotational phases at which the *Gaia* points were obtained. The common fit allows for interpolation of the maximum brightness from the fitted Fourier series. The interpolated maximum brightnesses for the *Gaia* points taken at different phase angles are then used in the construction of phase curves.

Similarly, the synthetic light curves generated from asteroid shape models can be composed with the *Gaia* data. This approach is, however, burdened with (1) uncertainty arising from propagating the uncertainty of the rotation period, and thus an inability to properly determine the rotational phase, and (2) light-curve amplitude uncertainties arising from shape models accuracy. This problem is minimized in our approach by the requirement for the same epoch for the dense photometry.

Generally, differential photometry can be re-calibrated to the *Gaia G* band, using one of the all-sky photometric catalogues. This, however, could possibly introduce unwanted systematic effects owing to the difference between photometric bands of *Gaia* and ground-based observations. These systematic effects are not well established at the moment and thus impossible to correct for.

In this work we focus on composing together the differential ground-based photometry with the *Gaia* absolute photometry. We outline our procedure below.

First, we use the PᴇʀFɪᴛ program (Kwiatkowski et al. 2021) to fit the standard Fourier series to the combined differential and absolute photometry. We determine the shape of the composite light curve and synodic period. We usually use 4–6 orders of the series; however, sometimes higher orders are required. The fitted function is

$$V(t) = \overline{V} + \sum_{k=1}^{n} \left( A_k \sin \frac{2\pi k(t-t_0)}{P} + B_k \cos \frac{2\pi k(t-t_0)}{P} \right), \quad (2)$$

where $\overline{V}$ is the average brightness for a single light curve, $A_k$ and $B_k$ are the Fourier series coefficients, $P$ is the synodic period, $t$ is the time of observation, and $t_0$ is the time of the first observation. For a fixed synodic period, that relationship is linear and can be fitted using a least-squares method. The fitted parameters become $\overline{V}$ (separate for each light curve), $A_k$ and $B_k$. A range of synodic periods is scanned to produce a plot of the quality of the fit (measured by $\chi^2$ per degree of freedom) versus the trial period and to find the best matching solution.

Provided that the data are relative or absolute in nature, the $\overline{V}$ plus half of the peak-to-peak amplitude (obtained from the Fourier fit) determines the maximum brightness ($V_{max}$). We thus further select only the $V_{max}$ originating from absolute *Gaia* data and use these to construct phase curves. Because *Gaia* observes asteroids mostly at phase angles away from opposition, only the linear part of the phase curve can be obtained. We thus use linear regression to fit the light-curve amplitude-corrected *Gaia* data. If relative photometry covering small phase angles is available (either dense or sparse from ground- or space-based observatories), full phase functions can be fitted. Therefore, this approach is capable of combining both dense and sparse data, relative, absolute and differential, providing a powerful tool for future-generation surveys.

In addition, because the largest uncertainties in correcting for light-curve amplitude arise from uncertainties in the synodic period, we perform an additional calibration step. That is, we sample the rotation period within the estimated uncertainty and select the period that leads to the smallest $\chi^2$ of the linear phase-curve fit.

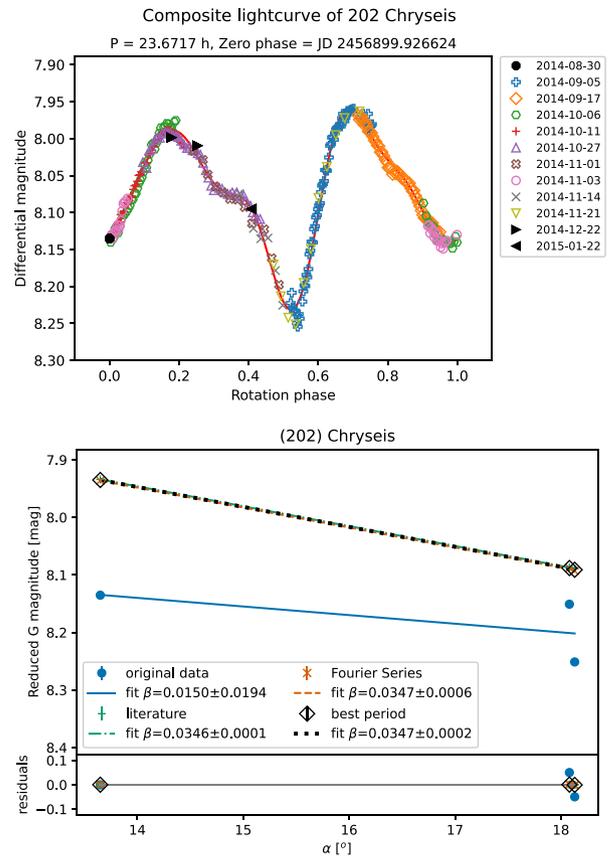

**Figure 1.** Upper panel: composite light curve for asteroid (202) Chryseis combining *Gaia* observations (black bold) and ground-based observations (coloured) with a derived light-curve shape (solid line). Bottom panel: linear part of the phase curve for asteroid (202) Chryseis. The original *Gaia* data reduced to 1-au distance and a linear fit to these are shown in blue. Light-curve amplitude-corrected *Gaia* data based on the literature, the derived and adjusted periods and corresponding fits are shown in green, orange and black, respectively. Error bars show the uncertainty of the brightness in magnitude. Residuals are shown at the very bottom.

## 4 RESULTS

### 4.1 *Gaia* data and differential ground-based photometry

We fitted the linear part of the phase curves to 32 different oppositions of 26 asteroids that met the selection criteria described in Section 2.

In Figs 1 to 3, we present composite light curves and final linear functions fitted to the light-curve-corrected data for asteroids (202) Chryseis, (301) Bavaria and (305) Gordonia (plots for the other asteroids can be found in the Supporting Information available online). The *Gaia* data nicely complement the ground-based dense differential light curves. LC-AC depends on the rotation period. The three linear functions shown were fitted to the data with LC-AC derived based on periods from Marciniak et al. (2015, 2018, 2019, 2021) and Shevchenko et al. (2021) (green), a period found using Fourier series (orange), and a period that minimizes the linear phase function residuals (black). The fit to the raw DR2 data is shown in blue. Generally, the LC-AC linear functions are similar, as the periods are usually consistent within hundredths of an hour. A linear fit to the raw *Gaia* data (without the LC-AC) clearly can produce erroneous and sometimes even negative (physically incorrect) phase slopes ($\beta$). Furthermore, the scatter around the linear functions is significantly reduced (e.g. from ∼0.15 to ∼0.02 mag in Fig. 3) compared with the linear function fitted to the unprocessed *Gaia* data. The reduction is larger for asteroids with high light-curve amplitude. This highlights the need for LC-AC when using sparse-in-time photometry. The reduction of the scatter is worse for asteroids with lower-quality light curves and for objects for which there is a gap in the coverage of the full rotational period. The linear LC-AC functions (Figs 1 to 3) correspond to light-curve maxima, and thus are usually shifted vertically compared with the raw *Gaia* data fit. The marked uncertainties in Figs 1 to 3 include only the single-point photometric uncertainty of the *Gaia* measurements and do not include the uncertainty arising from the propagation of synodic period uncertainty or synodic period changes.

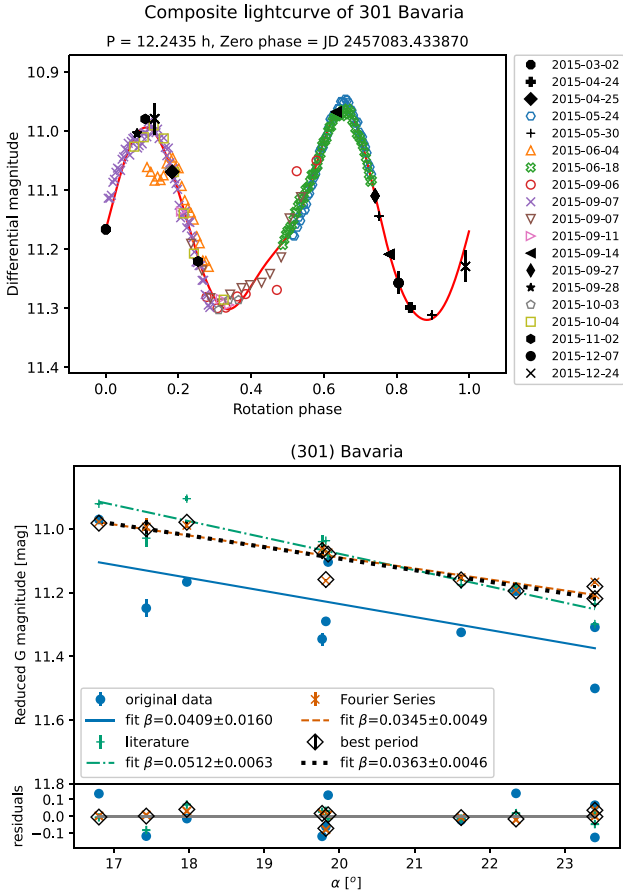

**Figure 2.** As Fig. 1, but for (301) Bavaria.

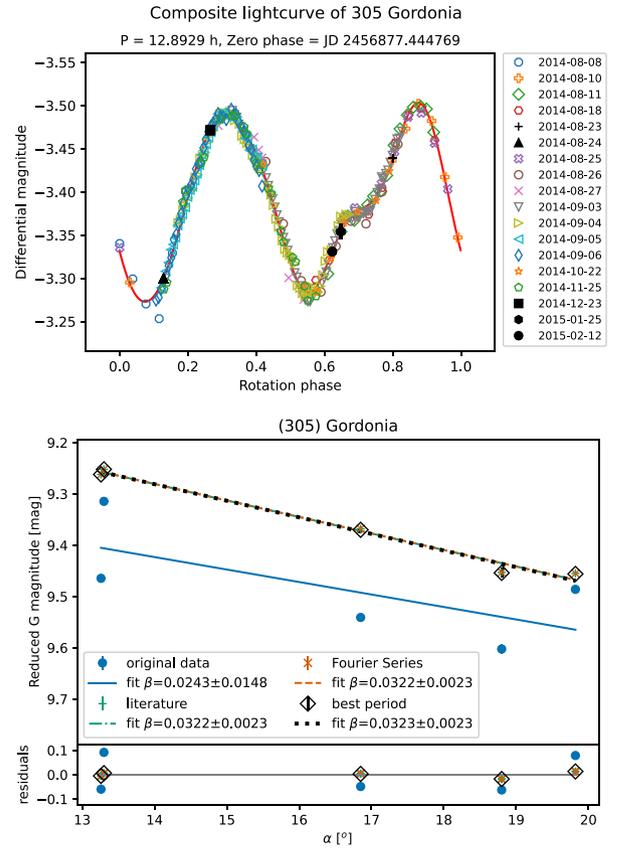

**Figure 3.** As Fig. 1, but for (305) Gordonia.

The resulting $\beta$ phase-slope values are listed in Table 2 along with the period that resulted in the best-fitting curve, Tholen taxonomic type (Neese 2010), and geometric albedos taken from the *AKARI* (Usui et al. 2011), *IRAS* (Tedesco et al. 2004), and *WISE* (Mainzer et al. 2019) catalogues. Data from different oppositions were fitted separately. The phase-slope values derived from different apparitions are typically within 1$\sigma$ uncertainty of each other. The differences may arise from different aspect angles and irregularities of the shape. Further uncertainties may originate from errors in photometry, period estimation and LC-AC.

Fig. 4 shows the correlation between the linear slope $\beta$ derived in this work and the geometric albedo. The solid black line represents the relationship $\beta = 0.016 - 0.022 \log p_{\rm V}$ found by Belskaya & Shevchenko (2018) for the *V*-band slopes, and the shaded areas represent 1$\sigma$, 2$\sigma$ and 3$\sigma$ error envelopes. The solid red line represents the function we fitted to the *Gaia G* magnitudes, which is different for every albedo source:

$$\beta_{\rm AKARI} = 0.025 - 0.013 \log p_{\rm V},$$
$$\beta_{\rm IRAS} = 0.022 - 0.015 \log p_{\rm V},$$
$$\beta_{\rm WISE} = 0.025 - 0.013 \log p_{\rm V}. \qquad (3)$$

We plot only points for which the accuracy is better than 20 per cent. It can be seen that our results are in good agreement with Belskaya & Shevchenko (2018). The relationships derived in equation (3) are for the *Gaia G* band. Owing to the sparse nature of the data, we consider these relationships to be rough estimates. Obtaining better relationships is not possible because it will probably never be possible to use the sparse data to improve the results based on accurate dense photometry.

### 4.2 Simulation of uncertainties

We checked how the time separation between dense and sparse observations (in assumption, differential and relative, respectively) and the noise of the dense observations affect the value and uncertainty of the derived slope parameter $\beta$. We selected the asteroid (159) Aemilia to perform our simulation, as it has the largest light-curve amplitude among our objects.

First, using observational data, we derived the mean JD time of ground-based data and set it to be the middle date of a week. Then we used ISAM to generate full synthetic light curves for three randomly chosen nights in a week and added 0.01-, 0.03- and 0.05-mag random Gaussian noise to them.

**Table 2.** Period used for composite light curves, geometric albedos from *AKARI* (Usui et al. 2011), *IRAS* (Tedesco et al. 2004), and *WISE* (Mainzer et al. 2019), linear slopes coefficients derived from the light-curve-corrected *Gaia* data, and Tholen taxonomic type from Neese (2010). In cases where there were a few entries for a given object in a given albedo catalogue, the entry with the lowest albedo uncertainty was taken.

| Asteroid | Period (h) | AKARI | $p_v$ IRAS | WISE | Phase slope [$\frac{mag}{deg}$] | Taxonomic type |
|---|---|---|---|---|---|---|
| (70) Panopaea | 15.8078 | 0.05 ± 0.002 | 0.07 ± 0.004 | 0.038 ± 0.007 | 0.0283 ± 0.0070 | C |
| (108) Hecuba | 14.2587 | 0.213 ± 0.007 | 0.192 ± 0.035 | 0.148 ± 0.013 | 0.0325 ± 0.0082 | S |
| (109) Felicitas | 13.1915 | 0.086 ± 0.003 | 0.06 ± 0.003 | 0.036 ± 0.009 | 0.0252 ± 0.0079 | GC |
| (159) Aemilia | 24.4822 | 0.059 ± 0.003 | 0.061 ± 0.003 | 0.064 ± 0.014 | 0.0382 ± 0.0042 | C |
| (195) Eurykleia | 16.5269 | 0.055 ± 0.002 | 0.053 ± 0.002 | 0.051 ± 0.008 | 0.0388 ± 0.0052 | C |
| (202) Chryseis | 23.6645 | 0.245 ± 0.007 | 0.178 ± 0.023 | 0.232 ± 0.037 | 0.0399 ± 0.0101 | S |
| (202) Chryseis | 23.6717 | 0.245 ± 0.007 | 0.178 ± 0.023 | 0.232 ± 0.037 | 0.0347 ± 0.0002 | S |
| (260) Huberta | 8.2883 | 0.054 ± 0.002 | 0.034 ± 0.005 | 0.044 ± 0.01 | 0.0408 ± 0.0186 | CX: |
| (301) Bavaria | 12.2435 | 0.06 ± 0.002 | 0.056 ± 0.003 | 0.057 ± 0.007 | 0.0363 ± 0.0046 | – |
| (305) Gordonia | 12.8929 | 0.234 ± 0.008 | 0.169 ± 0.01 | 0.238 ± 0.036 | 0.0323 ± 0.0023 | S |
| (329) Svea | 22.7646 | 0.049 ± 0.001 | 0.037 ± 0.002 | 0.039 ± 0.007 | 0.0117 ± 0.0143 | C |
| (362) Havnia | 16.9327 | 0.062 ± 0.002 | – | 0.061 ± 0.008 | 0.0308 ± 0.0134 | XC |
| (380) Fiducia | 13.7205 | 0.053 ± 0.002 | 0.051 ± 0.002 | 0.066 ± 0.005 | 0.0150 ± 0.0081 | C |
| (439) Ohio | 37.4550 | 0.037 ± 0.001 | 0.036 ± 0.002 | 0.042 ± 0.004 | 0.0430 ± 0.0078 | X: |
| (483) Seppina | 12.7219 | 0.172 ± 0.004 | 0.136 ± 0.007 | 0.184 ± 0.066 | 0.0401 ± 0.0103 | S |
| (483) Seppina | 12.7207 | 0.172 ± 0.004 | 0.136 ± 0.007 | 0.184 ± 0.066 | 0.0604 ± 0.0148 | S |
| (501) Urhixidur | 13.1738 | 0.079 ± 0.002 | 0.068 ± 0.004 | 0.052 ± 0.009 | 0.0399 ± 0.0051 | – |
| (537) Pauly | 16.3019 | 0.283 ± 0.008 | 0.239 ± 0.05 | 0.322 ± 0.03 | 0.0127 ± 0.0129 | DU: |
| (552) Sigelinde | 17.1532 | 0.051 ± 0.002 | 0.034 ± 0.003 | 0.036 ± 0.005 | 0.0438 ± 0.0025 | – |
| (611) Valeria | 6.9788 | 0.115 ± 0.003 | 0.091 ± 0.005 | 0.124 ± 0.008 | 0.0362 ± 0.0050 | S |
| (618) Elfriede | 14.7980 | 0.06 ± 0.002 | 0.058 ± 0.007 | 0.05 ± 0.005 | 0.0583 ± 0.0174 | C |
| (618) Elfriede | 14.7974 | 0.06 ± 0.002 | 0.058 ± 0.007 | 0.05 ± 0.005 | 0.0463 ± 0.0039 | C |
| (653) Berenike | 12.4876 | 0.173 ± 0.006 | 0.177 ± 0.011 | 0.085 ± 0.017 | 0.0858 ± 0.0435 | S |
| (666) Desdemona | 14.7686 | 0.105 ± 0.006 | 0.095 ± 0.005 | 0.095 ± 0.015 | 0.0638 ± 0.0375 | – |
| (666) Desdemona | 14.6148 | 0.105 ± 0.006 | 0.095 ± 0.005 | 0.095 ± 0.015 | 0.0353 ± 0.0020 | – |
| (667) Denise | 12.6840 | 0.062 ± 0.003 | 0.057 ± 0.003 | 0.062 ± 0.01 | 0.0401 ± 0.0037 | – |
| (667) Denise | 12.6840 | 0.062 ± 0.003 | 0.057 ± 0.003 | 0.062 ± 0.01 | 0.0455 ± 0.0037 | – |
| (723) Hammonia | 5.4351 | 0.294 ± 0.031 | 0.121 ± 0.017 | 0.352 ± 0.048 | 0.0145 ± 0.0052 | – |
| (727) Nipponia | 5.0692 | 0.212 ± 0.01 | 0.141 ± 0.018 | 0.479 ± 0.053 | 0.0863 ± 0.0175 | DT |
| (780) Armenia | 19.8799 | 0.046 ± 0.002 | 0.047 ± 0.002 | 0.039 ± 0.003 | 0.0423 ± 0.0114 | – |
| (834) Burnhamia | 13.8753 | 0.082 ± 0.007 | 0.068 ± 0.004 | 0.071 ± 0.008 | 0.0403 ± 0.0035 | GS: |
| (834) Burnhamia | 13.8764 | 0.082 ± 0.007 | 0.068 ± 0.004 | 0.071 ± 0.008 | 0.0252 ± 0.0061 | GS: |

Next, to simulate *Gaia* observations, for three randomly selected days in each of the following 10 weeks, we generated a synthetic light curve, picked a single observing point, and assigned a sigma equal to 0.003 (the median of all *Gaia* points sigma). Finally, we created 30 different data sets containing all possible combinations of differently noised synthetic ground-based data mixed with simulated *Gaia* points from 1 to 10 weeks apart from dense ground-based ones. We performed the calibration and phase-curve fitting described in Section 3 on all data sets. The obtained light curves and phase curves can be found in the Supporting Information available online. Fig. 5 shows the dependence of the $\beta$ parameter and its uncertainty on the time span between observations.

First, the dependence of the data-separation time-span on the slope uncertainty is non-linear and relies on the number of rotations in the separation time span, the period uncertainty, and the phase difference created by the difference in the synodic and sidereal periods and the rotational phase of the sparse data. In addition, based on this single simulation, we recommend that the dense observations should not have $\sigma$ values larger than 0.03 mag.

## 5 DISCUSSION

The largest source of uncertainty in phase-curve studies based on sparse data arises from the lack of or an inaccurate LC-AC,

as the sparse data randomly sample light curves of generally unknown shape, rotation period and amplitude. Therefore phase curves constructed from sparse photometry commonly have a large scatter around the fitted phase function. Naturally, the scatter is larger for objects with a large light-curve amplitude, which for some objects can reach even $>\sim 1$ mag (Warner, Harris & Pravec 2009). Absolute magnitudes derived from the scattered phase curves typically correspond to light-curve-averaged brightnesses. Moreover, the average is taken from points that are not uniformly sampled over the rotation period but depend on survey cadence, and this may introduce biases in the derived phase-curve parameters.

For example, Oszkiewicz et al. (2011) and Mahlke et al. (2021) did not correct for light-curve amplitude for the majority of their data. The estimated absolute magnitudes from Oszkiewicz et al. (2011) thus correspond to brightness averages over randomly sampled light curves from multiple oppositions, and those by Mahlke et al. (2021) to randomly sampled light curves over a single opposition. Waszczak et al. (2015) simultaneously fitted phase curves together with the rotational component (using second-order Fourier series fits representing simple ellipsoidal shapes) using sparse photometry from the Palomar Transient Factory. Vereš et al. (2015) used sinusoidal fits to describe light curves composed of sparse data and to determine the maximum brightness. That approach does not take into account the changes in synodic period or complicated light-curve morphology.

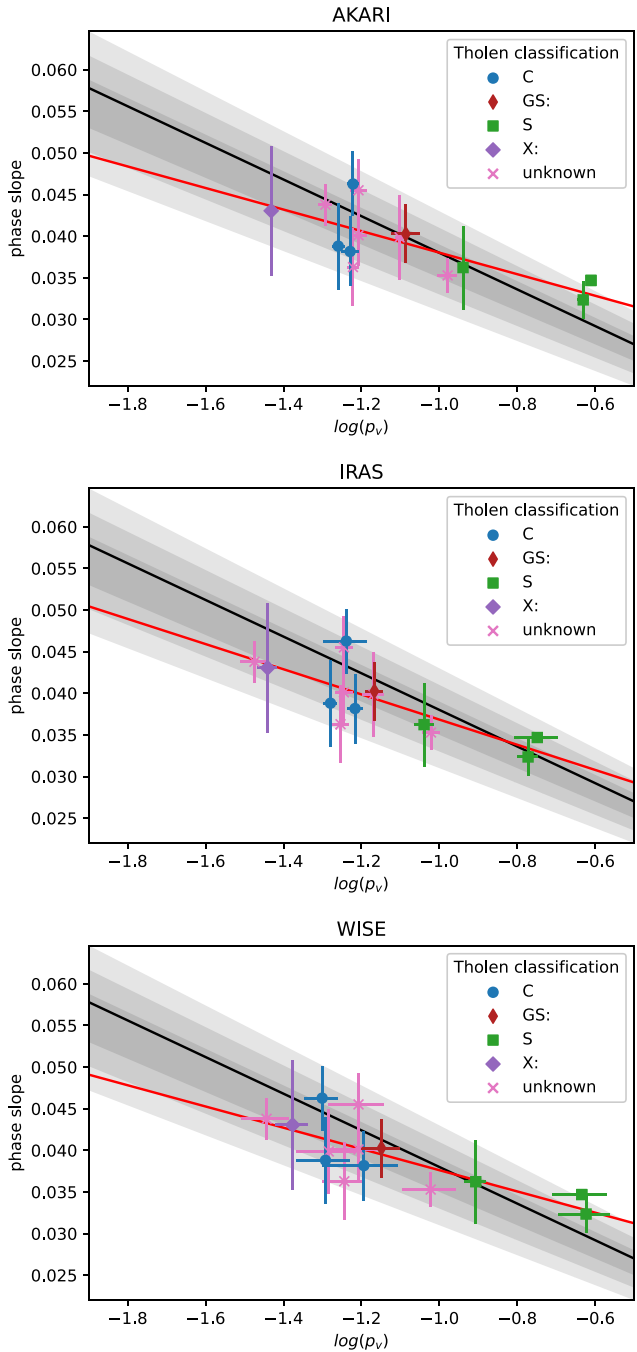

**Figure 4.** Correlation between the slope of asteroid phase curves and the albedo. The black line with the envelope depicts $\beta = 0.016 - 0.022\log p_v$ with $1\sigma$, $2\sigma$ and $3\sigma$ (Belskaya & Shevchenko 2018), the red line shows a function fitted to the data points, and the points are objects from this study.

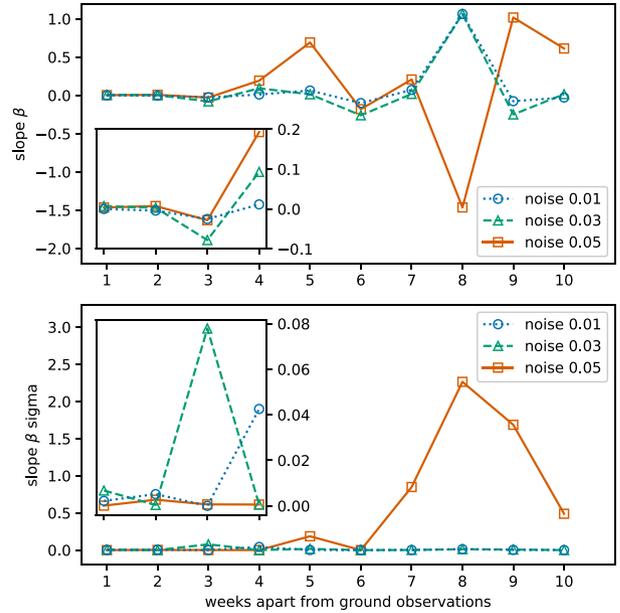

**Figure 5.** Data-separation time-span dependence for synthetic observations of asteroid (159) Aemilia. The plot shows changes of the slope value $\beta$ (upper) and its uncertainty (bottom) depending on the time-distance of sparse observations. The inset graphs are close-up views of the range 1–4 weeks.

Alvarez-Candal et al. (2021a) used Bayesian statistics to account for possible light-curve amplitudes and periods. Their study did not consider aspect changes in different oppositions and was based on a few data points per object, resulting in large uncertainties. Colazo et al. (2021) combined mean magnitudes (averaged over photometric measurements obtained by *Gaia*, given that there was more than one transit per Julian day) with ground-based relative *V*-band photometry from Lagerkvist et al. (1995). The ground-based data were obtained at different epochs and thus at different aspects compared with the *Gaia* data. In the *Gaia* DR2 documentation, the brightest measurements among photometric points obtained on a single night are selected for phase-curve analysis. In general, approximate light-curve correction methods could lead to a larger spread in points around the phase function and thus to higher phase-curve parameter uncertainties.

Muinonen et al. (2020) and Martikainen et al. (2020) used a complex convex inversion method that allows for accurate LC-AC. However, if no dense light curves are available around the sparse data, the correction may similarly be inaccurate owing to model uncertainties. Moreover, if the shapes are derived based on sparse photometry they may not accurately reflect the light-curve morphology at *Gaia* epochs. Furthermore, because most asteroids have a poorly constrained shape and dimension along the *z*-axis, the transformation to the equatorial viewing geometry applied by the authors may introduce additional uncertainties (Bartczak & Dudziński 2019).

In this work, we have presented an alternative to the methods mentioned above. The approach presented here is straightforward, computationally efficient, and has not yet been explored in the literature. Our approach allows for the incorporation of differential photometry, which traditionally is unusable for the construction of phase curves. The requirement of dense photometry close in time to the sparse measurements reduces the spread of data around the fitted phase function. In comparison to other works, this allows us to account for the complicated shape of the light-curve morphology beyond the second order of Fourier series, as often simplified in previous works. Because phase curves are derived from data from single oppositions, the resulting absolute magnitudes and phase-curve parameters correspond to that opposition.

This method should be used for asteroids for which the aspect changes are slow, which is generally the case for main-belt objects. As a precaution, we show a phase curve derived for asteroid (109) Felicitas in Fig. 6, which had a large aspect change of 40°.

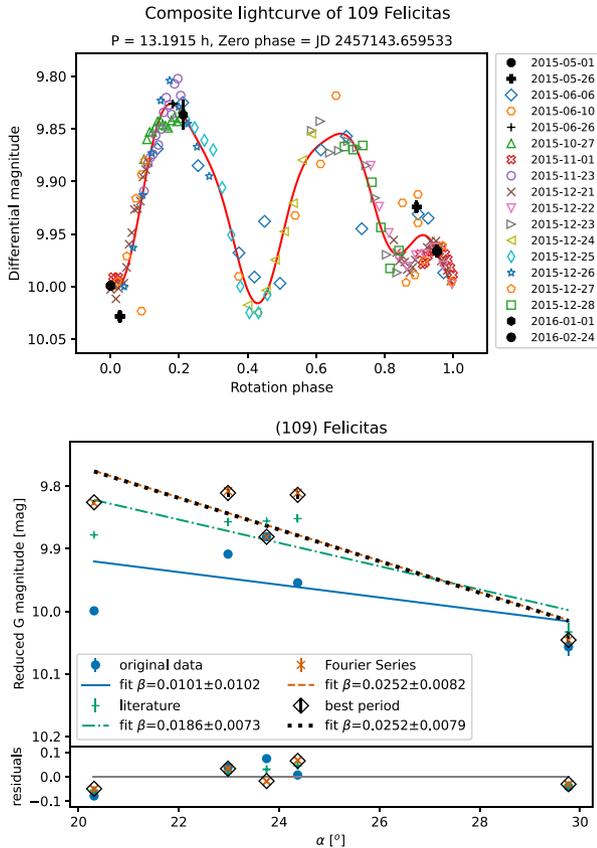

**Figure 6.** As Fig. 1, but for (109) Felicitas.

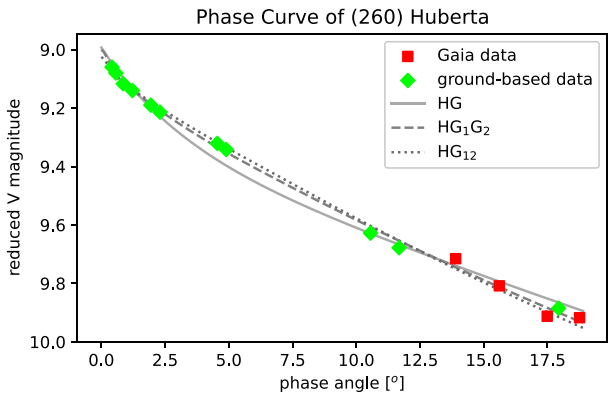

**Figure 7.** Full phase curve of (260) Huberta with fitted $H, G$, $H, G_1, G_2$ and $H, G_{12}$ phase functions (solid, dashed and dotted lines, respectively).

**Table 3.** Phase-function parameters for (260) Huberta.

| $H, G$ | $H = 8.99^{+0.02}_{-0.03}$ |
| | $G = 0.20^{+0.02}_{-0.05}$ |
| $H, G_1, G_2$ | $H = 9.00^{+0.02}_{-0.03}$ |
| | $G_1 = 0.69^{+0.08}_{-0.09}$ |
| | $G_2 = 0.12^{+0.05}_{-0.04}$ |
| $H, G_{12}$ | $H = 9.01^{+0.01}_{-0.01}$ |
| | $G_{12} = 0.66^{+0.13}_{-0.01}$ |

taken from Shevchenko et al. (2021). The obtained phase function parameters are summarized in Table 3.

Generally, having accurate models and data close in time to the sparse data would be best for reducing the phase-curve uncertainties. Our method, by reproducing the complicated light-curve shape, reduces the uncertainty due to rotation and shape. However, it still cannot reproduce the changes in the synodic period, which can be obtained from models. Both spin/shape models and Fourier fits suffer from the propagation of synodic period uncertainty. The rotational phase uncertainty increases linearly with time (Bartczak & Dudziński 2019). Thus, composing sparse data with dense light curves that have been obtained at a very different epoch may result in a large rotational phase discrepancy and thus induce additional errors in the interpolation of maximum brightness. Thus, having data very close in time to the sparse photometry will best reduce these sources of uncertainty.

Aspect changes also influence phase curves. Traditionally, phase curves are obtained for objects with small aspect changes (most main-belt objects) to avoid the uncertainties induced by aspect. The method of Martikainen et al. (2021) proposed computing so-called reference phase curves, namely phase curves in the equatorial geometry (often referred to as aspect-corrected). These, however, suffer from additional uncertainties arising from the shape models, which typically have large uncertainties in the $z$-axis. The use of dense absolute and relative photometry could help with this problem. These types of data can help us to constrain the axial ratio, and thus also improve the $z$-axis uncertainty.

## 6 SUMMARY AND FUTURE WORK

Because the number of photometric measurements for asteroids is growing exponentially in time, it is important to develop methods capable of combining these distributed data. In this work, we have presented an efficient method that is able to produce phase curves from combined dense and sparse photometry, and the rotational brightness modulation is accounted for. Moreover, we have shown that differential photometry (unusable for the construction of traditional phase curves) can be utilized with sparse data, aiding the light-curve correction. Furthermore, increasing the number of sparse data will allow for the combination of different data sets into 'denser' light curves, which in turn will provide a higher quality of phase curves.

In the future, we plan to develop a large photometric data base combining sparse and dense photometry from various publicly available data sources. This will include absolute, relative, and differential photometry.

Combining a larger amount of data will allow for a more accurate determination of phase function parameters for a large number of asteroids. This in turn will lead to a better determination of asteroid

There is almost no reduction of the original scatter around the linear phase curve.

In addition, we would like to point out that by combining calibrated ground-based observations with sparse data, a full phase curve can be obtained and phase-function parameters can be determined. As an example, we created a phase curve for the object (260) Hammonia combining ground-based $V$-filter data from Shevchenko et al. (2021) and data from *Gaia* DR2 (Fig. 7). *Gaia* magnitudes have been transformed to $V$ Johnson–Cousins magnitudes using the transformation from Busso et al. (2018) and the colour index $V$–$R$ sizes, albedos, and other surface properties that are correlated with phase-curve parameters. Furthermore, poorly understood problems such as phase-curve parameters relation to taxonomy, wavelength dependence, and aspect angle dependence can be studied given a large data base containing improved phase curves.

## ACKNOWLEDGEMENTS

This work has been supported by grant no. 2017/25/B/ST9/00740 from the National Science Centre, Poland. VS and IB gratefully acknowledge support from the National Research Foundation of Ukraine (grant no. 2020.02/0371). DO was supported by the National Science Centre, Poland, grant no. 2017/26/D/ST9/00240. This work has made use of data from the European Space Agency (ESA) mission *Gaia* (https://www.cosmos.esa.int/gaia), processed by the *Gaia* Data Processing and Analysis Consortium (DPAC, https://www.cosmos.esa.int/web/gaia/dpac/consortium). Funding for the DPAC has been provided by national institutions, in particular the institutions participating in the *Gaia* Multilateral Agreement.

## DATA AVAILABILITY

The majority of the photometric light curves are already available in the CDS (Centre de Données astronomiques de Strasbourg) archive. The remaining unpublished light curves will be submitted to CDS. Composite light curves and phase curves for the other objects in this work as well as for the simulated data are available in the online Supporting Information.

## SUPPORTING INFORMATION

Supplementary data are available at *MNRAS* online.

Please note: Oxford University Press is not responsible for the content or functionality of any supporting materials supplied by the authors. Any queries (other than missing material) should be directed to the corresponding author for the article.

## APPENDIX A: OBSERVATIONAL DETAILS

**Table A1.** Observing circumstances (date of observation, ecliptic longitude $\lambda$, ecliptic latitude $\beta$, phase angle $\alpha$, observer and observing site) for the new ground-based light curves of (611) Valeria and (727) Nipponia used in this paper.

| Date | $\lambda$ (°) | $\beta$ (°) | $\alpha$ (°) | Observer | Site |
|---|---|---|---|---|---|
| (611) Valeria | | | | | |
| 2014-10-04 | 22.61 | −3.21 | 4.34 | K. Sobkowiak | Borowiec, Poland |
| 2014-10-24 | 18.07 | −4.78 | 4.99 | A. Marciniak | Borowiec, Poland |
| | | | | | |
| (727) Nipponia | | | | | |
| 2015-02-17 | 140.83 | 4.58 | 3.52 | J. Horbowicz | Borowiec, Poland |
| 2015-02-18 | 140.78 | 4.60 | 3.60 | V. Kudak, V. Perig | Derenivka, Ukraine |
| 2015-03-17 | 135.60 | 6.61 | 14.77 | P. Kankiewicz | Kielce, Poland |
| 2015-03-19 | 135.42 | 6.71 | 15.38 | A. Marciniak | Borowiec, Poland |
| 2015-04-15 | 135.68 | 7.63 | 21.02 | P. Kulczak | Borowiec, Poland |

This paper has been typeset from a TeX/LaTeX file prepared by the author.

## SUPPLEMENTARY FIGURES

The graphs shown in Fig. 1 to 28 and in Table 1 consist of two parts: upper and bottom. Upper part is a composite lightcurve for a given asteroid combining Gaia observations (black bold) and ground-based observations (color) with a derived lightcurve shape (solid red line). Bottom part shows a linear part of the phase curve. Original Gaia data reduced to 1 au distance and a linear fit to those is shown in blue. Lightcurve amplitude corrected Gaia data based on literature, derived and adjusted period and corresponding fits are shown in green, orange and black respectively. Error bars show uncertainty of brightness in magnitude. Residuals are shown at the very bottom.

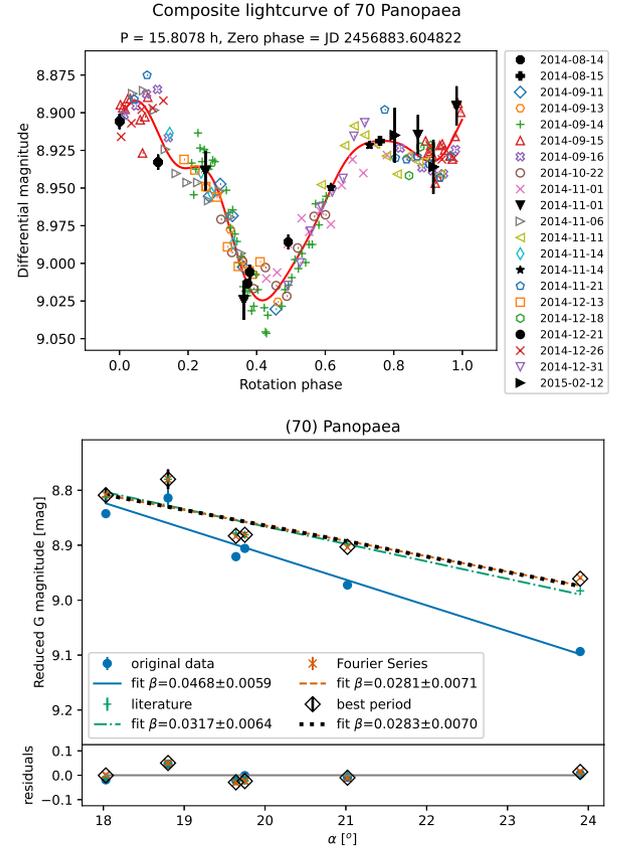

**Figure 1.** (70) Panopaea

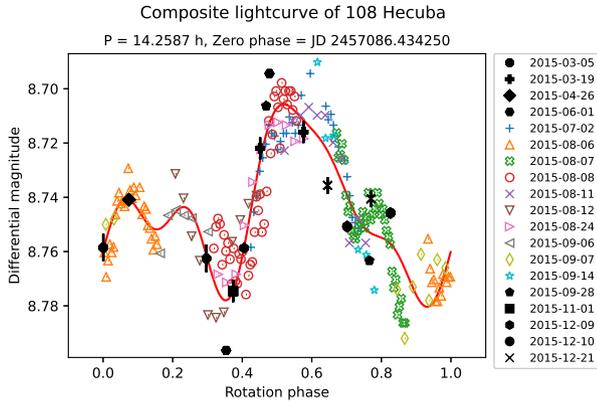

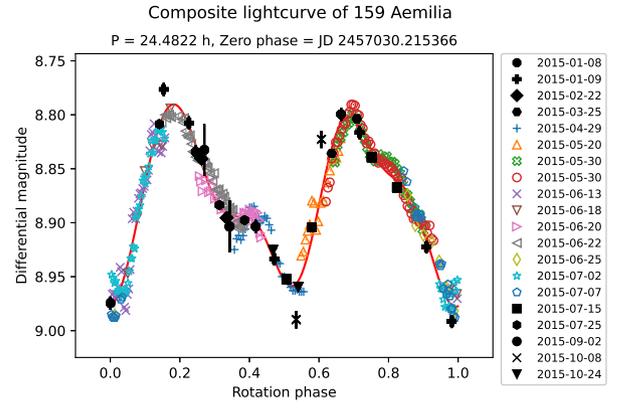

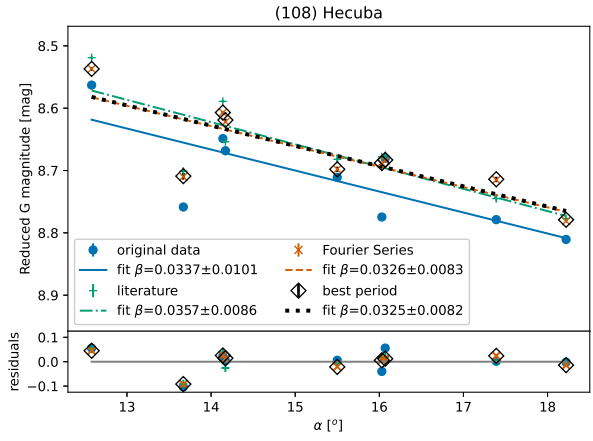

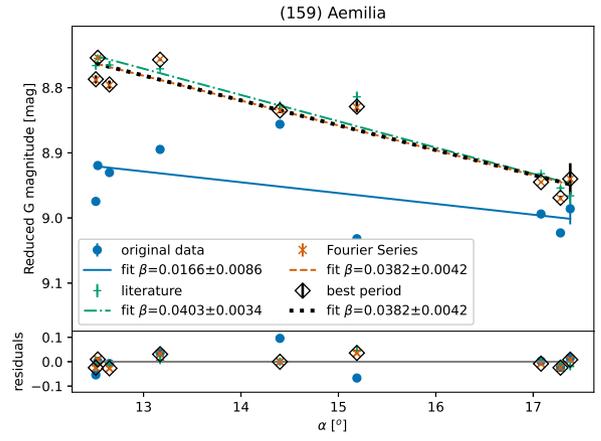

**Figure 2.** (108) Hecuba

**Figure 3.** (159) Aemilia

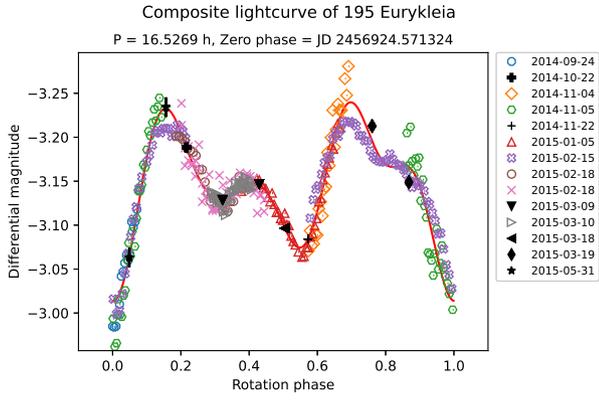

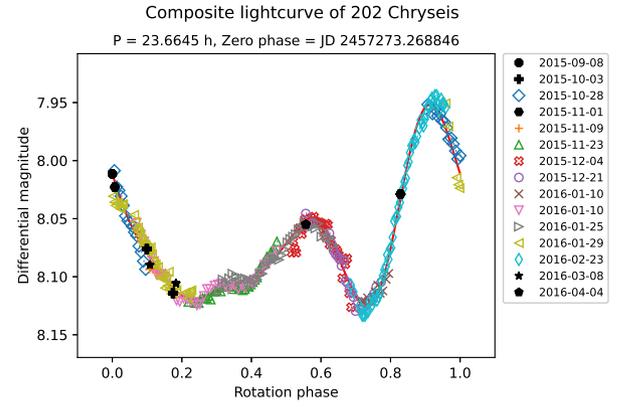

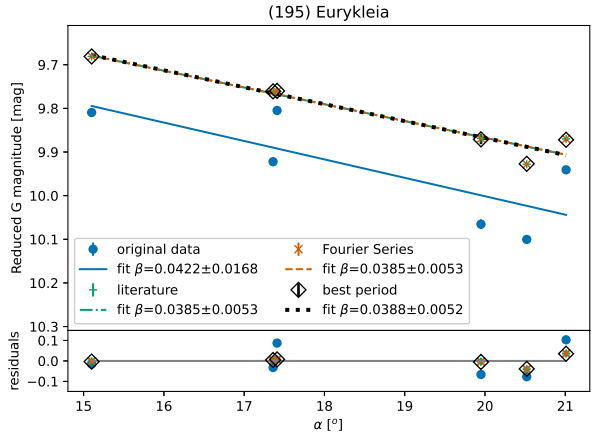

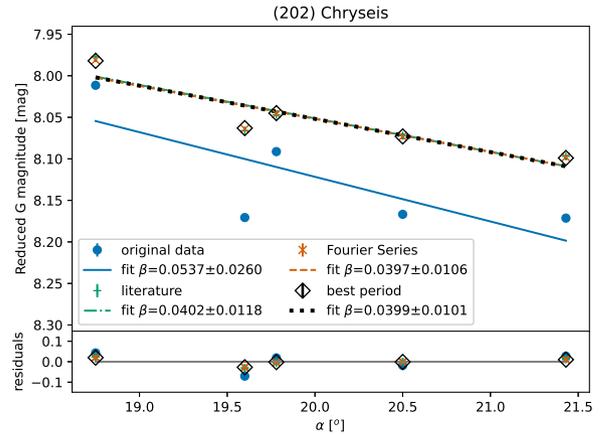

**Figure 4.** (195) Eurykleia

**Figure 5.** (202) Chryseis

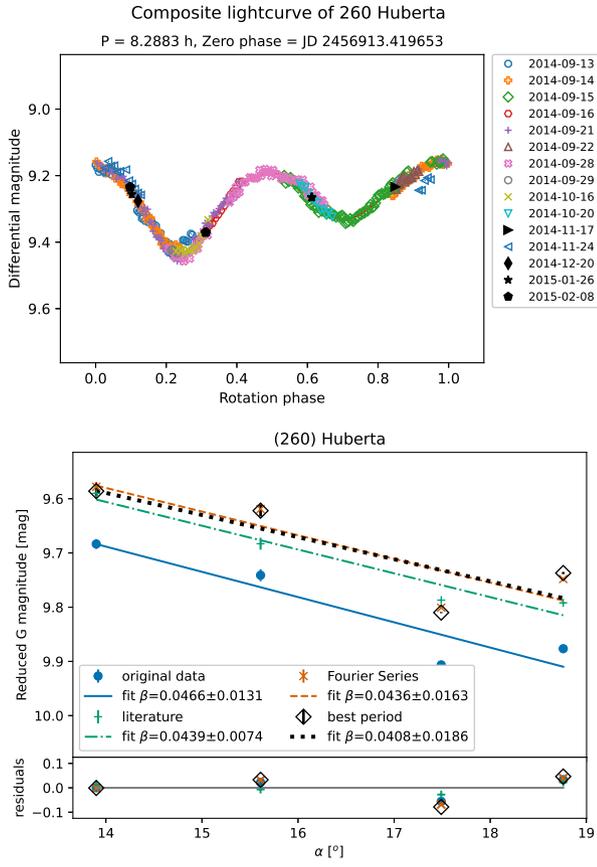

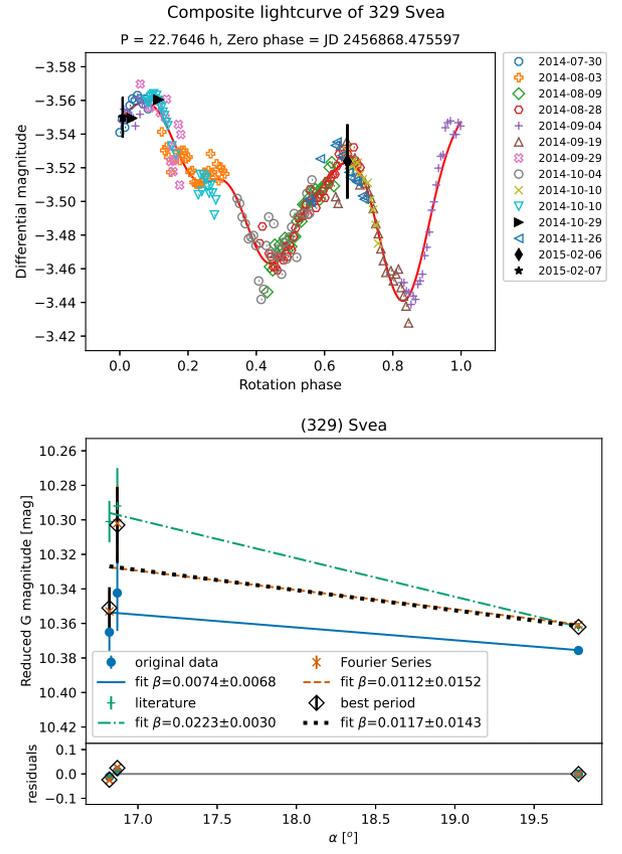

**Figure 6.** (260) Huberta

**Figure 7.** (329) Svea

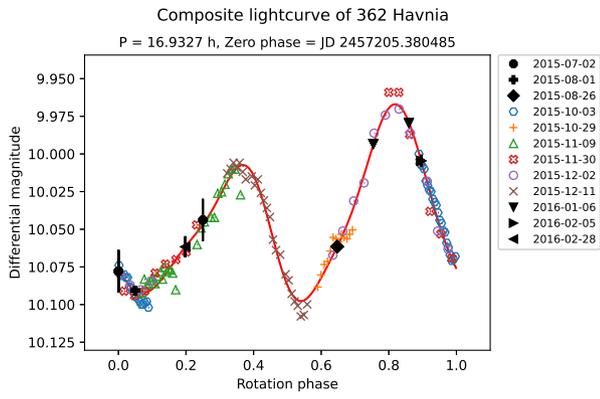

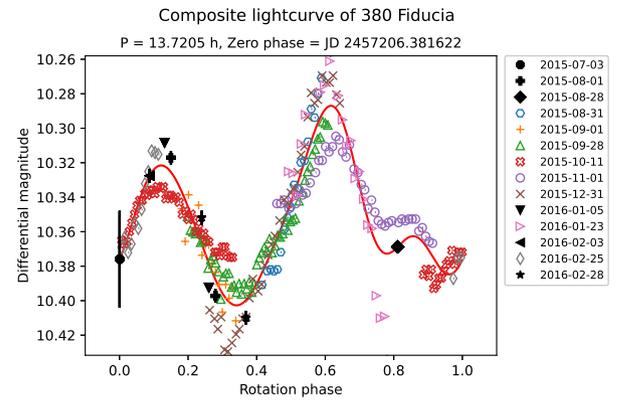

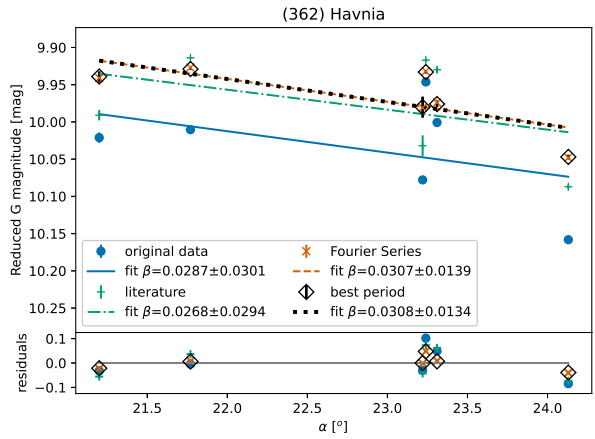

**Figure 8.** (362) Havnia

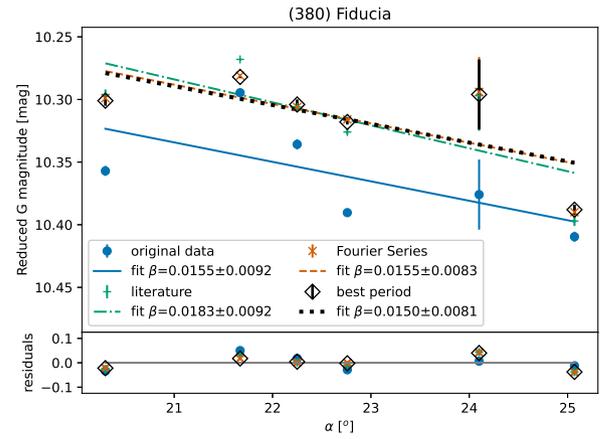

**Figure 9.** (380) Fiducia

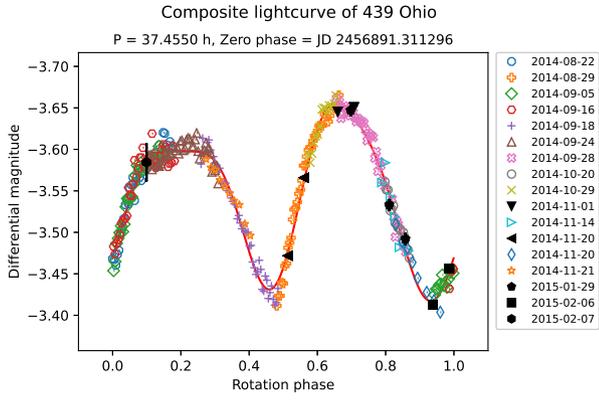

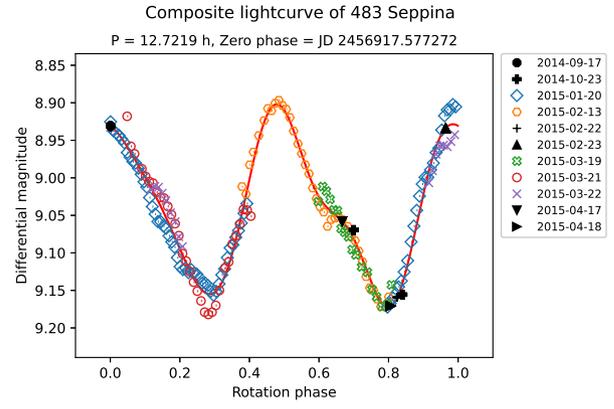

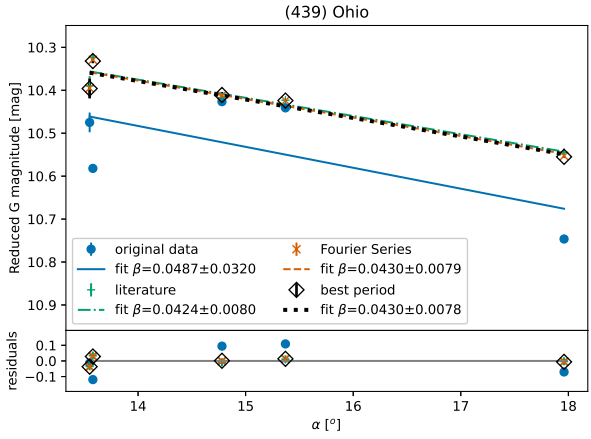

**Figure 10.** (439) Ohio

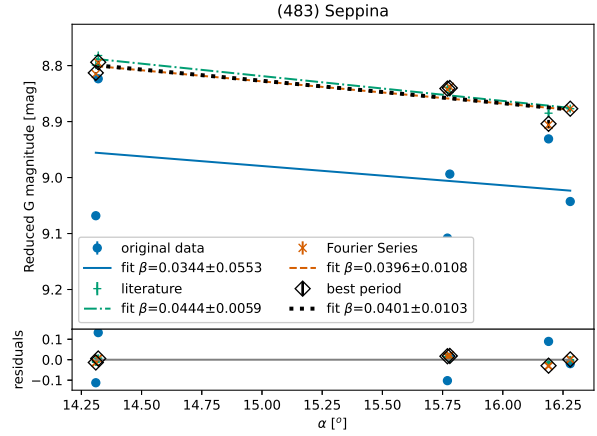

**Figure 11.** (483) Seppina

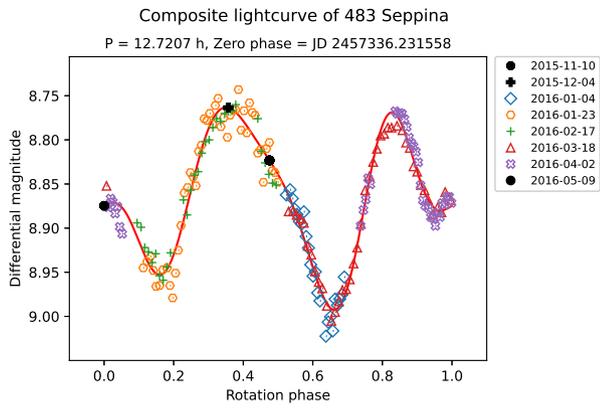

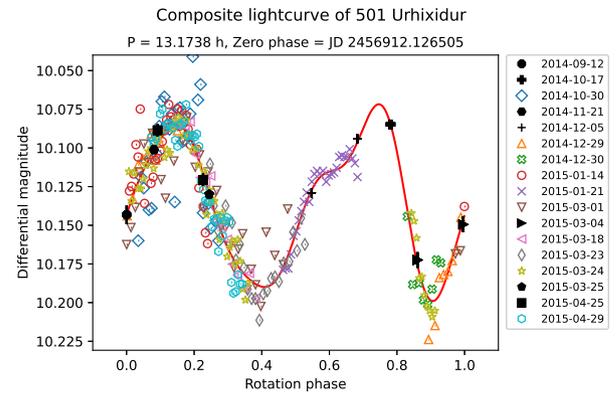

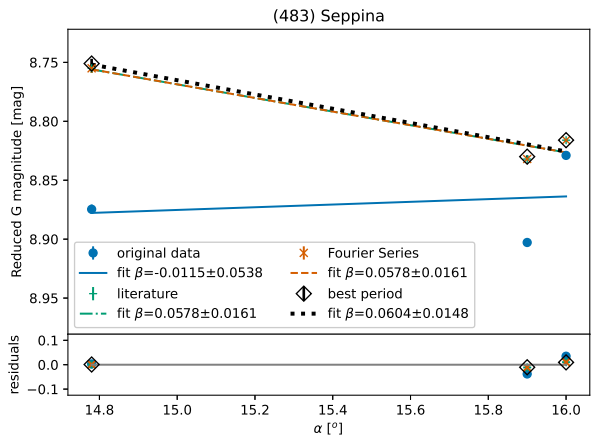

**Figure 12.** (483) Seppina

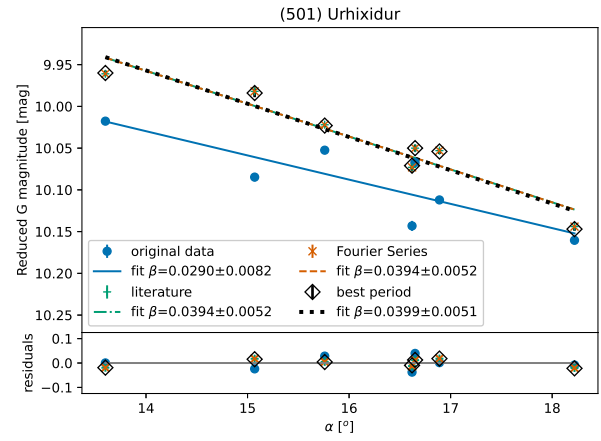

**Figure 13.** (501) Urhixidur

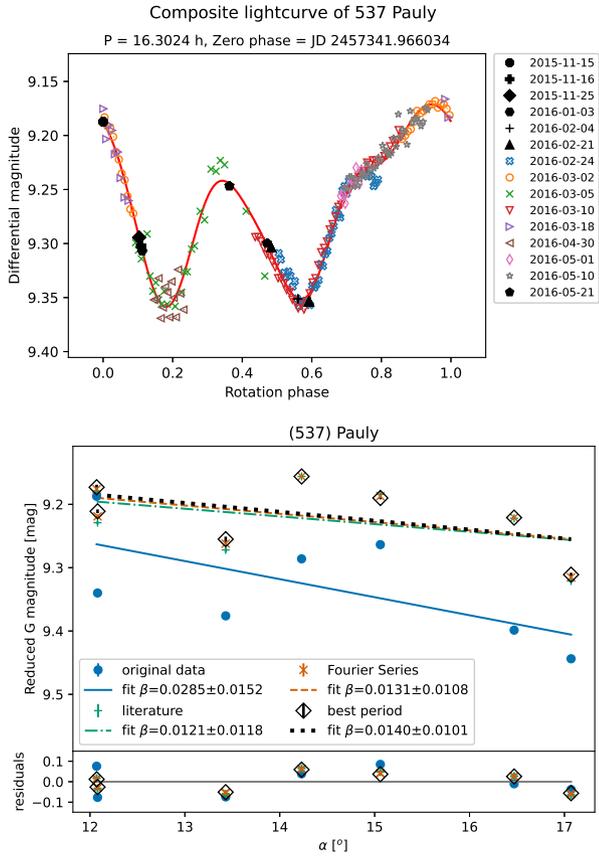

**Figure 14.** (537) Pauly

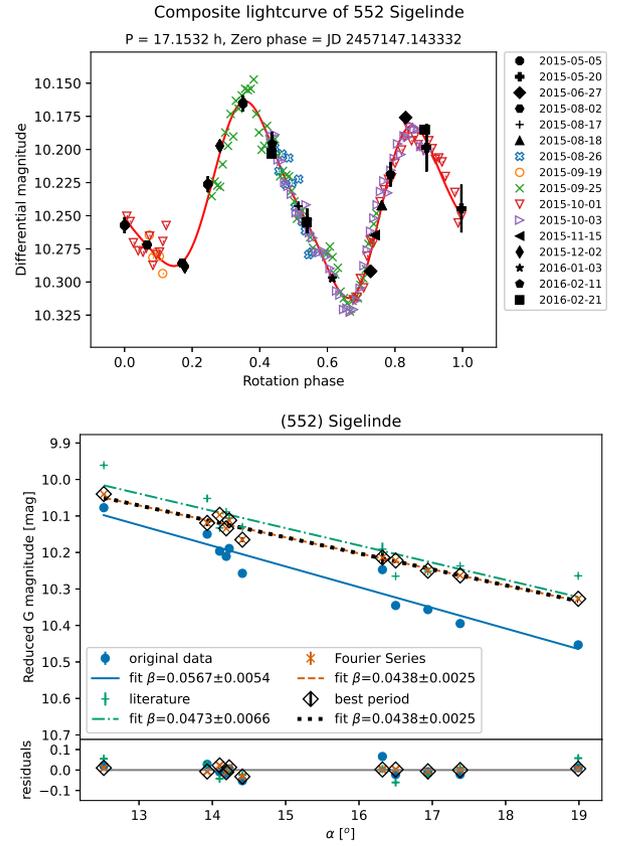

**Figure 15.** (552) Sigelinde

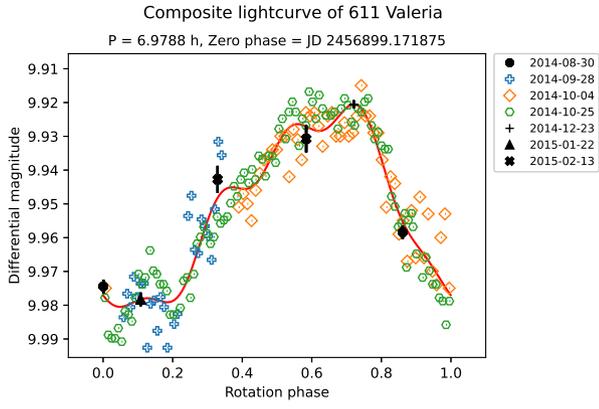

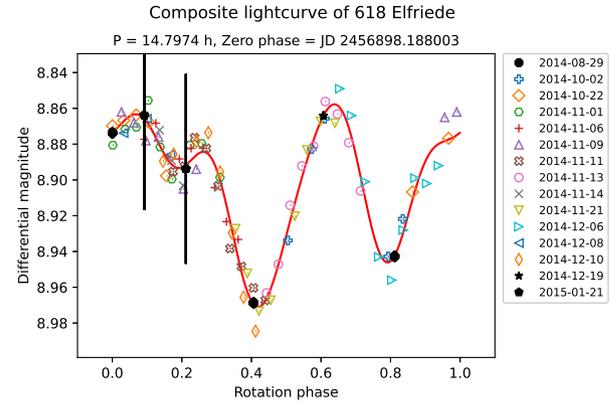

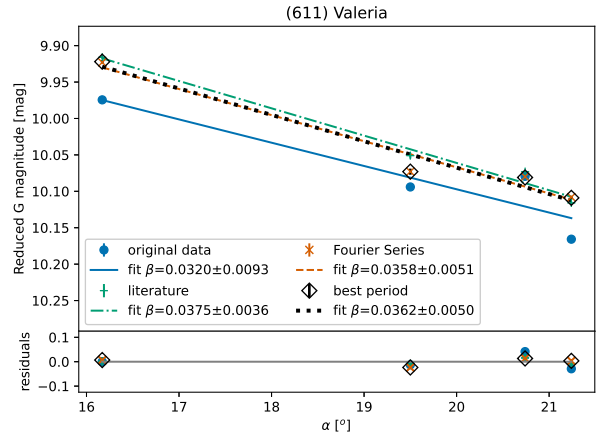

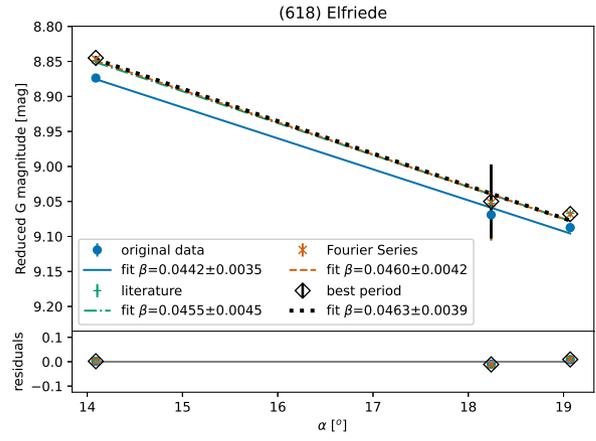

**Figure 16.** (611) Valeria

**Figure 17.** (618) Elfriede

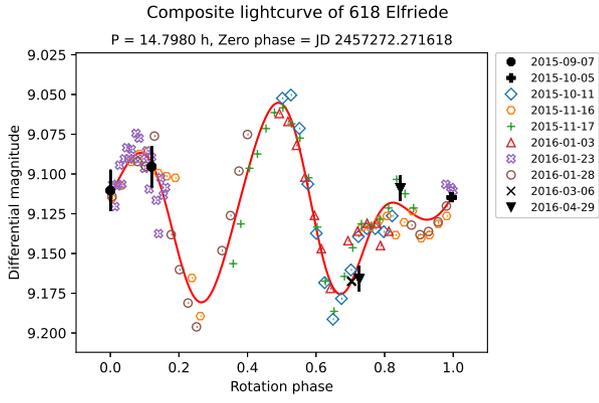

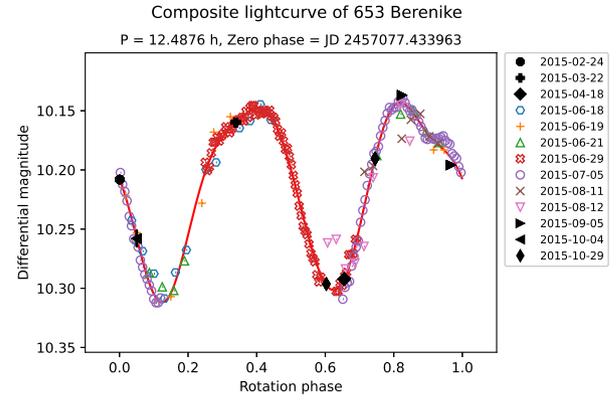

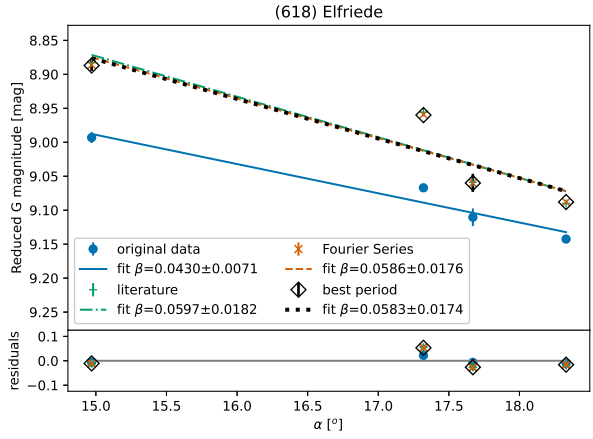

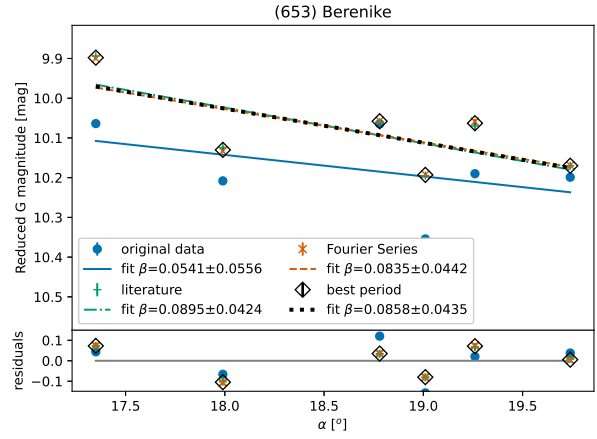

**Figure 18.** (618) Elfriede

**Figure 19.** (653) Berenike

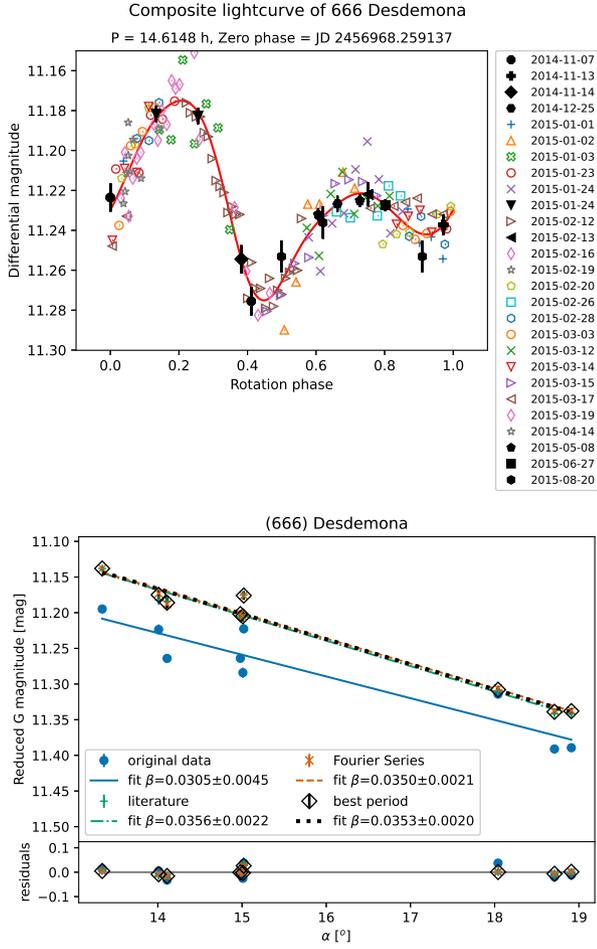

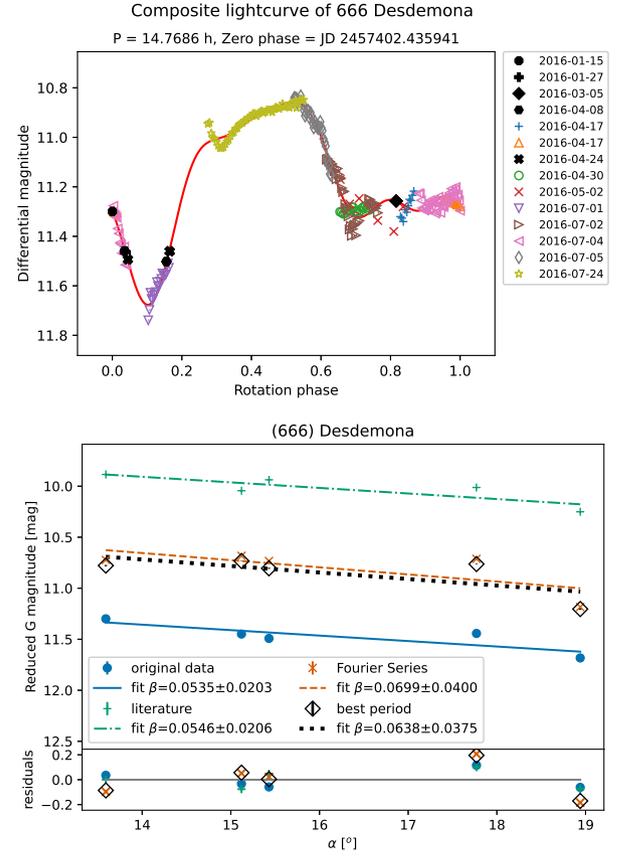

**Figure 20.** (666) Desdemona

**Figure 21.** (666) Desdemona

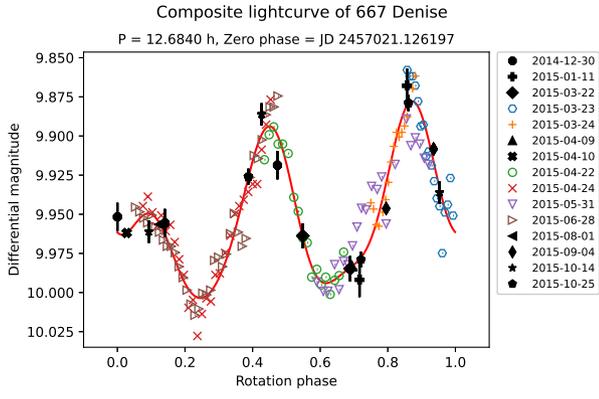

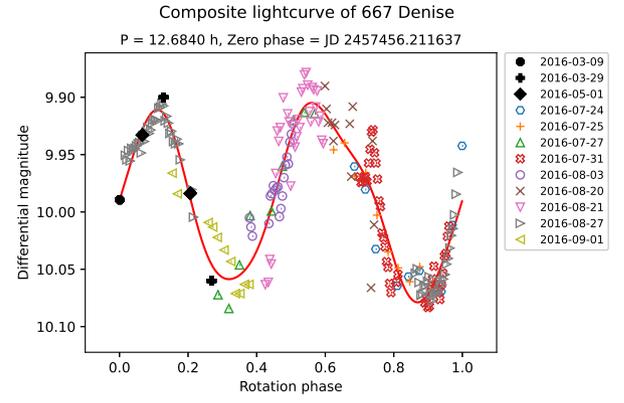

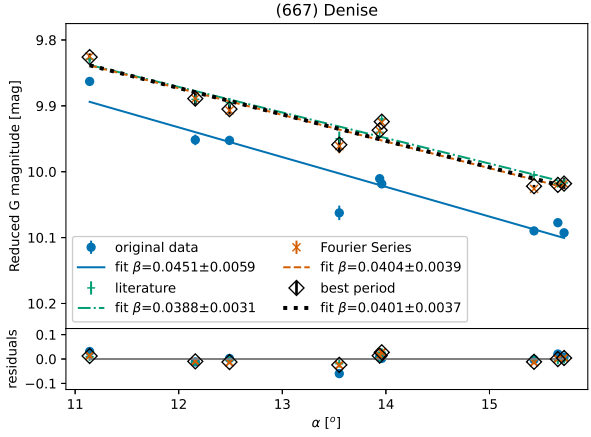

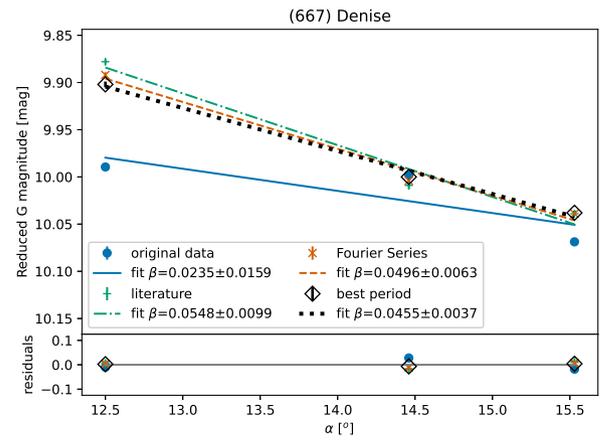

**Figure 22.** (667) Denise

**Figure 23.** (667) Denise

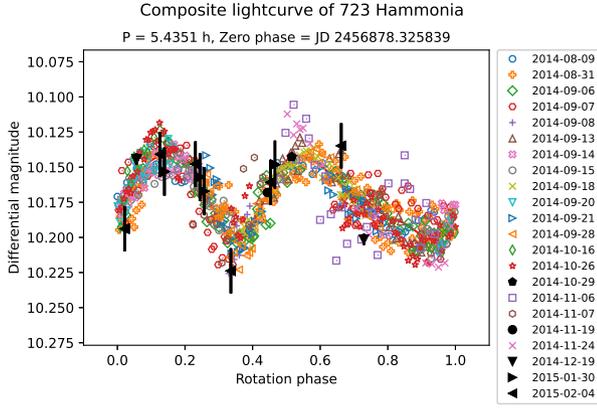

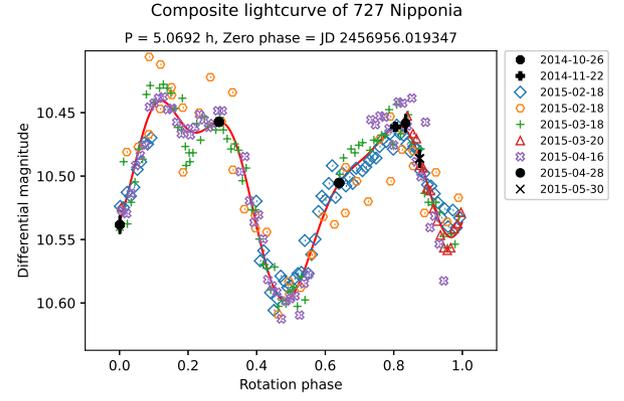

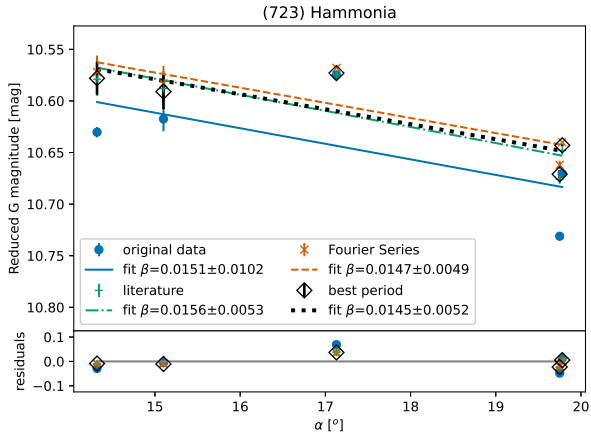

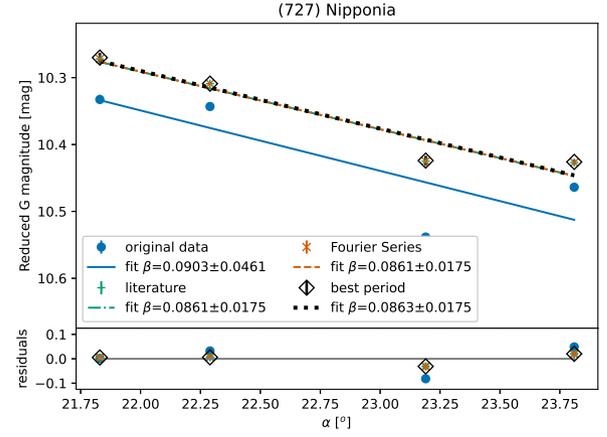

**Figure 24.** (723) Hammonia

**Figure 25.** (727) Nipponia

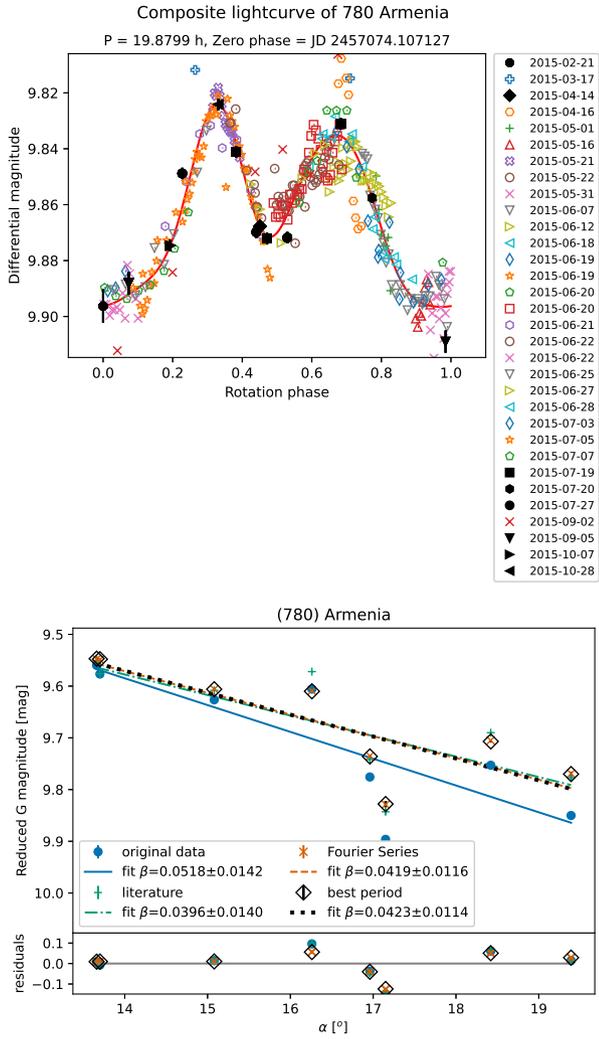

**Figure 26.** (780) Armenia

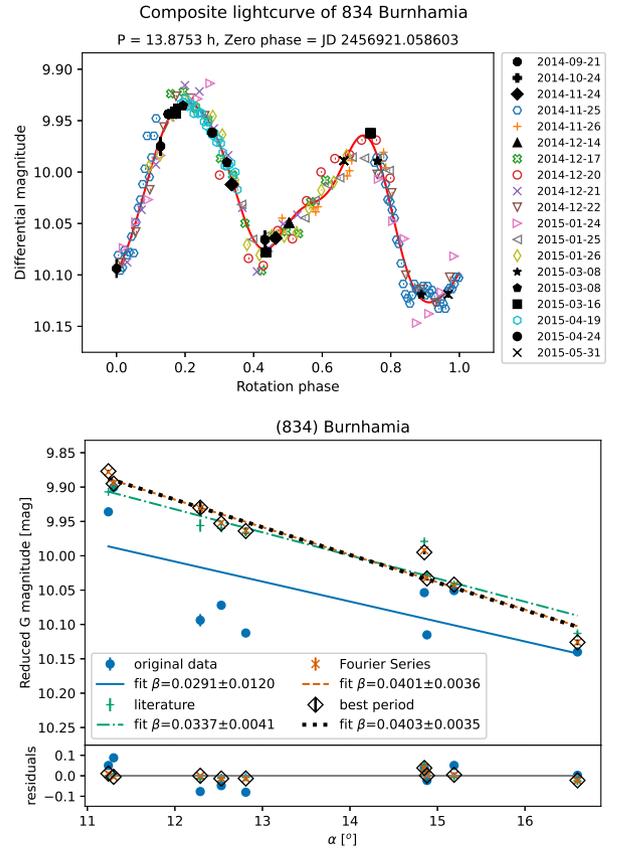

**Figure 27.** (834) Burnhamia

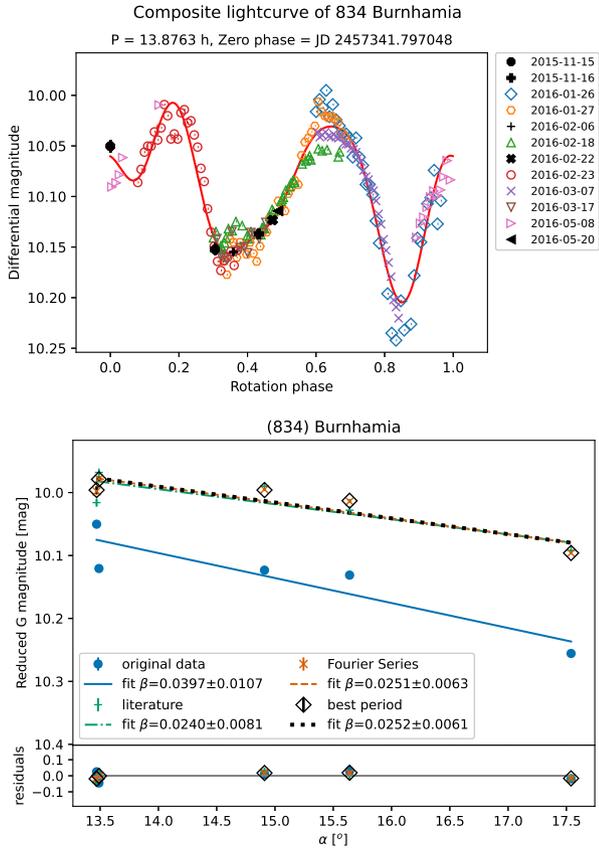

**Figure 28.** (834) Burnhamia

Table 1: Set of figures same as Fig. 1 showing how the time separation (rows) between dense and sparse observations (in assumption, differential and relative, respectively) and the noise (columns) of the dense observations affects lightcurves and phase curves for simulated data of (159) Aemilia.

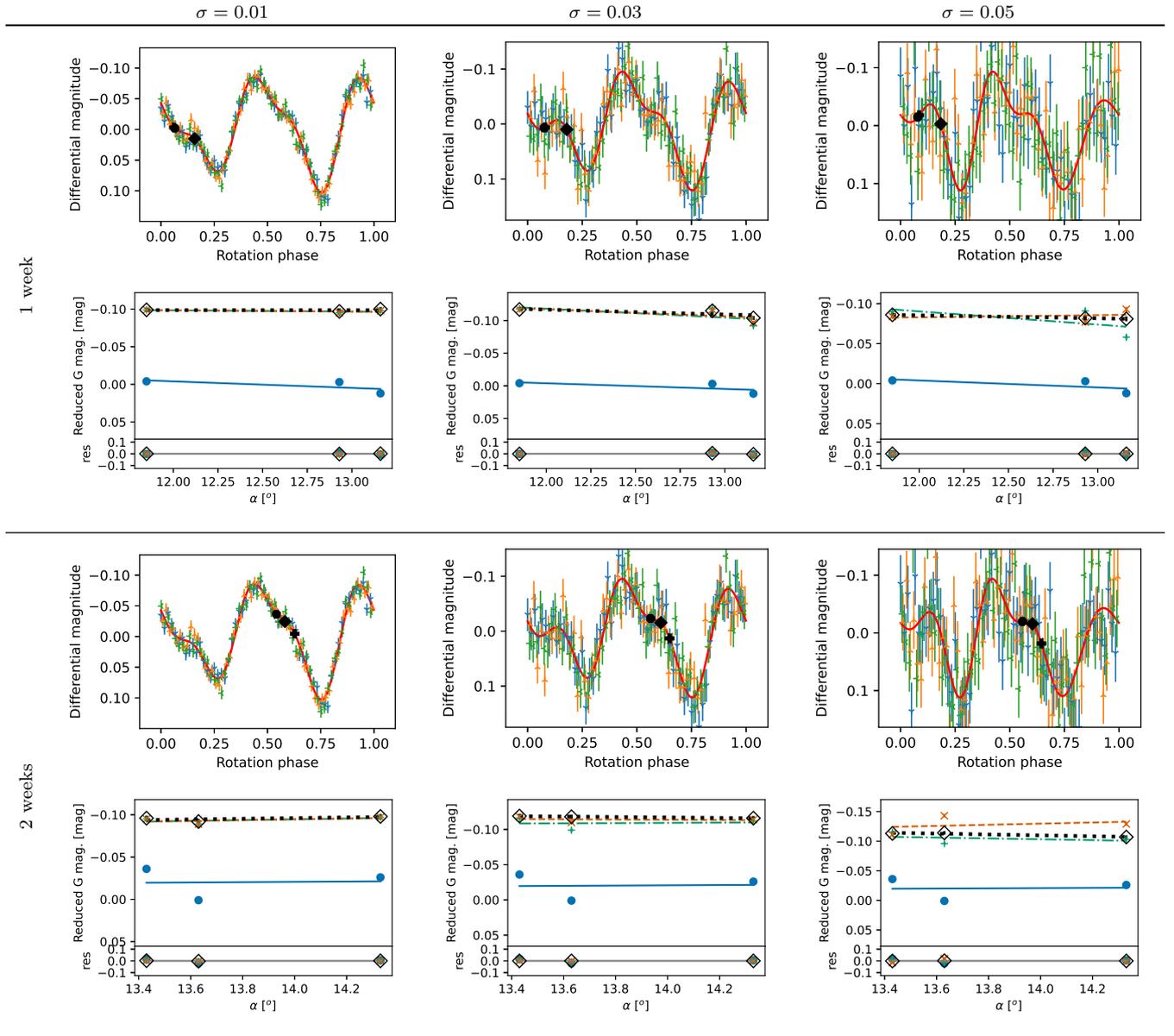

Table 1: Set of figures same as Fig. 1 showing how the time separation (rows) between dense and sparse observations (in assumption, differential and relative, respectively) and the noise (columns) of the dense observations affects lightcurves and phase curves for simulated data of (159) Aemilia.

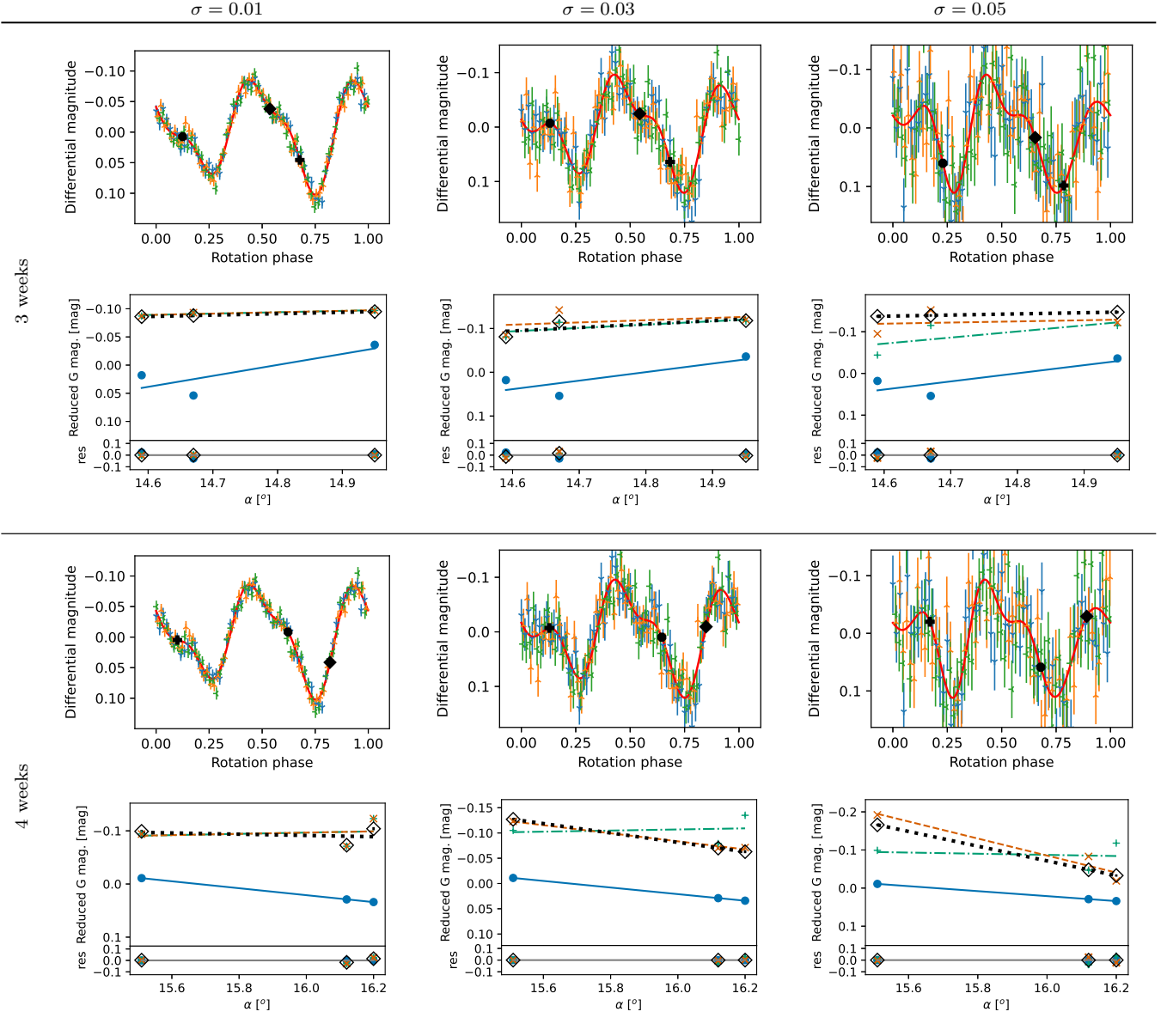

Table 1: Set of figures same as Fig. 1 showing how the time separation (rows) between dense and sparse observations (in assumption, differential and relative, respectively) and the noise (columns) of the dense observations affects lightcurves and phase curves for simulated data of (159) Aemilia.

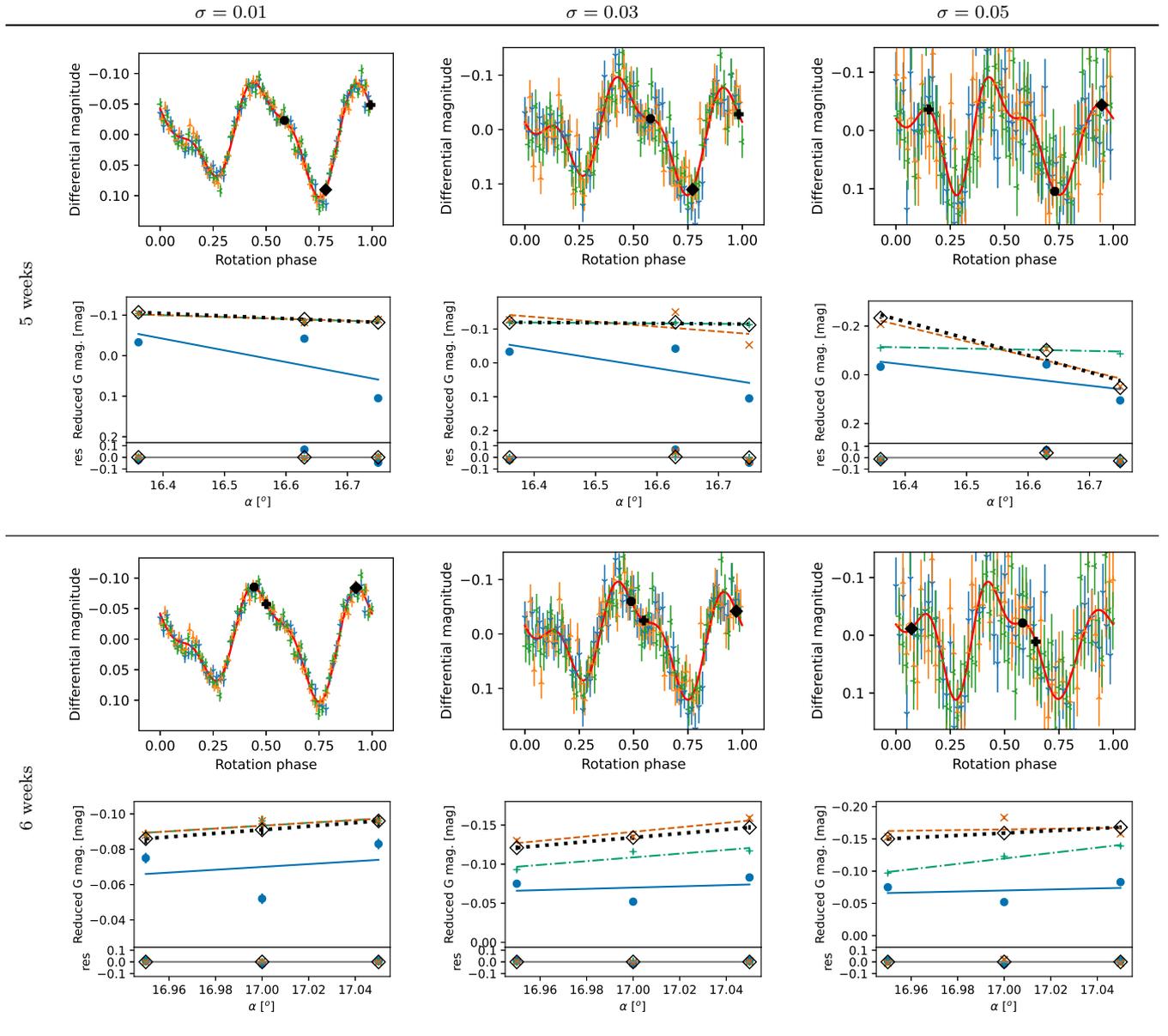

Table 1: Set of figures same as Fig. 1 showing how the time separation (rows) between dense and sparse observations (in assumption, differential and relative, respectively) and the noise (columns) of the dense observations affects lightcurves and phase curves for simulated data of (159) Aemilia.

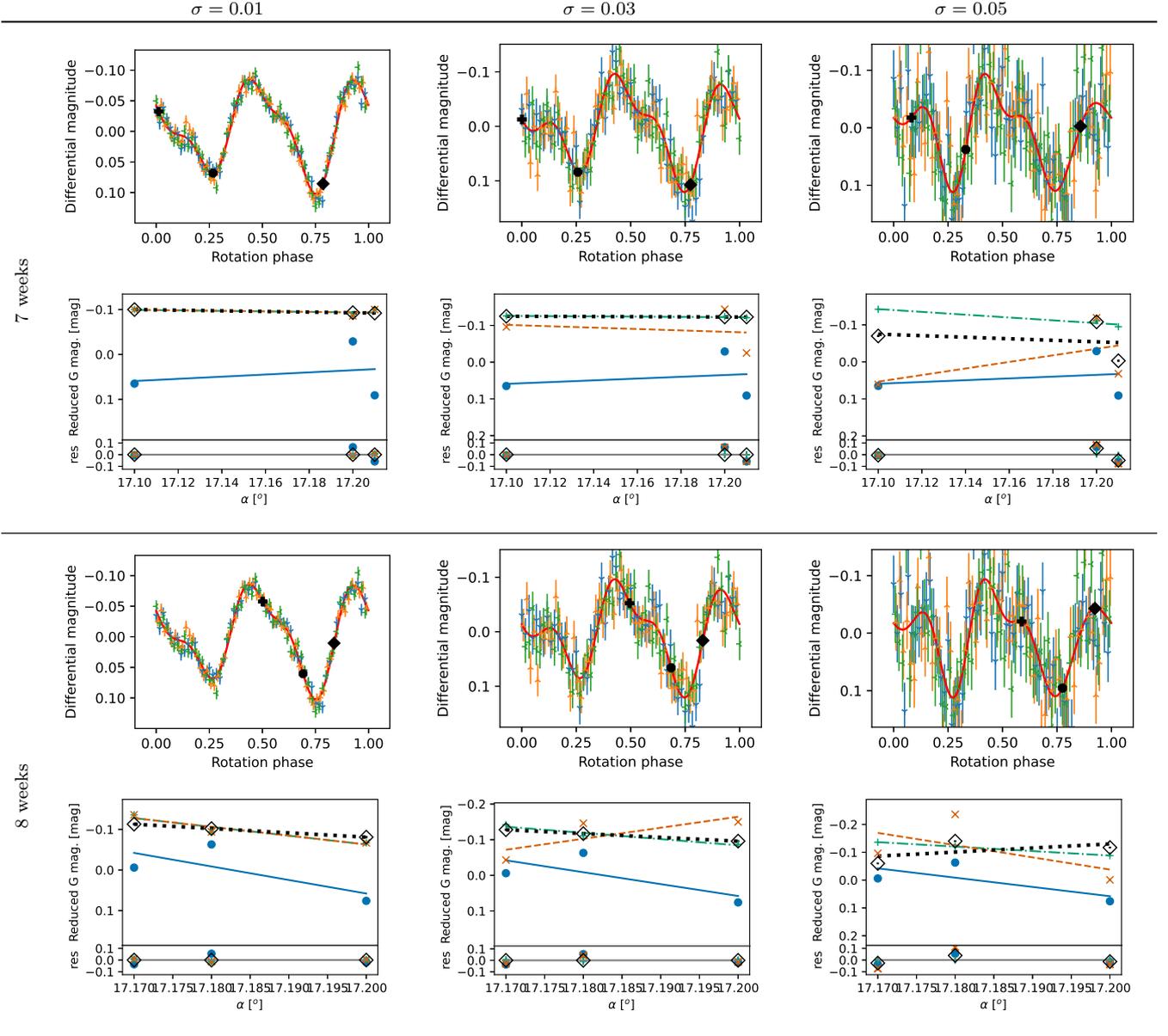

Table 1: Set of figures same as Fig. 1 showing how the time separation (rows) between dense and sparse observations (in assumption, differential and relative, respectively) and the noise (columns) of the dense observations affects lightcurves and phase curves for simulated data of (159) Aemilia.

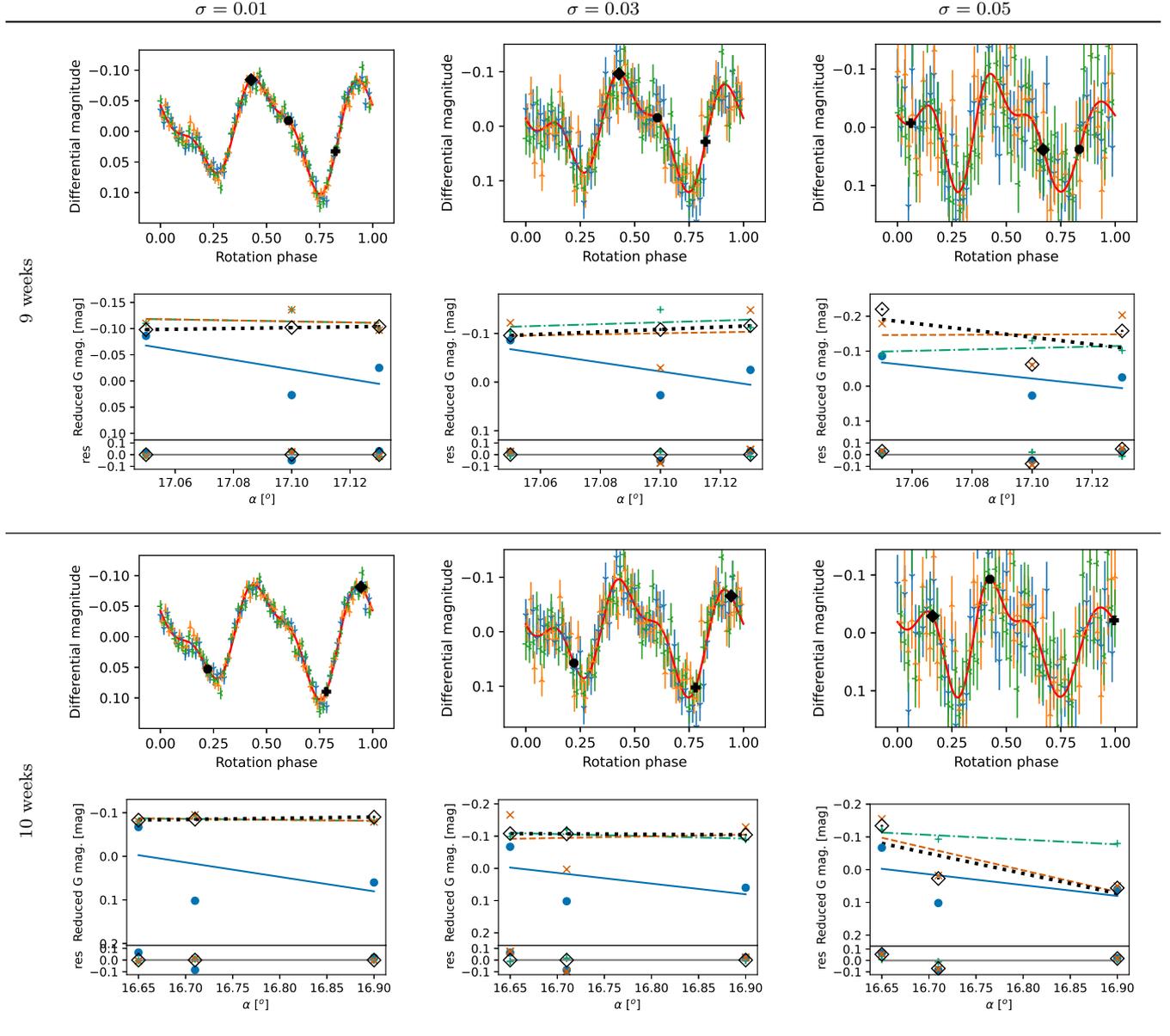